\documentclass[12pt]{article}
\pdfoutput=1

\usepackage{amsmath,amssymb,graphicx} 
\usepackage{epsf}
\usepackage{pstricks}
\usepackage{cite}
\usepackage{multirow}

\newcommand{\beq}{\begin{eqnarray}}
\newcommand{\eeq}{\end{eqnarray}}

\newcommand{\centeron}[2]{{\setbox0=\hbox{#1}\setbox1=\hbox{#2}\ifdim
                                        
\wd1>\wd0\kern.5\wd1\kern-.5\wd0\fi
\copy0

\kern-.5\wd0\kern-.5\wd1\copy1\ifdim\wd0>\wd1
                                       \kern.5\wd0\kern-.5\wd1\fi}}
\newcommand{\ltap}{\>\centeron{\raise.35ex\hbox{$<$}}
                               {\lower.65ex\hbox{$\sim$}}\>}
\newcommand{\gtap}{\>\centeron{\raise.35ex\hbox{$>$}}
                               {\lower.65ex\hbox{$\sim$}}\>}

\newcommand\ZZ{\hbox{\zfont Z\kern-.4emZ}}
\font\zfont = cmss10 

\begin{document}
\begin{titlepage}
\begin{flushright}
LYCEN 2012-06
\end{flushright}

\vskip.5cm
\begin{center}
{\huge \bf 
Even tiers and resonances \\
\vspace{0.5cm}
on the Real Projective Plane.
}

\vskip.1cm
\end{center}
\vskip0.2cm

\begin{center}
{\bf
{Giacomo Cacciapaglia}, {\rm and}
{Bogna Kubik}}
\end{center}
\vskip 8pt

\begin{center}
{\it Universit\'e de Lyon, F-69622 Lyon, France; Universit\'e Lyon 1, Villeurbanne;\\
CNRS/IN2P3, UMR5822, Institut de Physique Nucl\'eaire de Lyon\\
F-69622 Villeurbanne Cedex, France } \\
\end{center}

\vglue 0.3truecm

\begin{abstract}
\vskip 3pt
\noindent
In this work we focus on various phenomenological aspects of the lightest even tiers, $(2,0)$ and $(0,2)$, in models based on a Real Projective Plane in 6 dimensions. 
We discuss the spectrum of the levels due to loop corrections, and the limit when the two radii are equal, in which case the two levels mix with each other and a new basis is defined. 
We also discuss the dependence of the spectrum on the ratio of the two radii.
These results are essential to understand the phenomenology of the model at colliders (LHC) and to predict the relic abundance of Dark Matter. 
Finally, we estimate the bounds on the radius from resonant decays of the even tiers at the LHC, showing that they can be in the $600$ GeV range after the complete analysis of the 2011 data.

\end{abstract}

\end{titlepage}

\newpage


\section{Introduction}
\label{sec:intro0}
\setcounter{equation}{0}
\setcounter{footnote}{0}

We live in exciting times for particle physics! Since its first physics run, the Large Hadron Collider (LHC) experiments have been collecting a considerable amount of high quality data.
The results so far have been disappointing for Beyond the Standard Model (BSM) physics, because the first searches have accurately confirmed the expectations for a Standard Model (SM) only hypothesis, and no trace of new physics have appeared.
Many models, like simple or constrained supersymmetric models, Randall-Sundrum gravitons, large extra dimensions ({\it \`a la} ADD), TeV-scale Black Holes, excited fermions and leptoquarks, fourth generation of fermions, have been disfavoured or severely constrained.
However, this was the {\it easy catch}, selected for an early discovery, and many simple and motivated models are still poorly constrained by the present data.
The ATLAS and CMS collaborations have also announced the discovery of a new resonance at a mass of $125\div126$ GeV, that has similar properties as the SM Higgs boson, thus confirming the hints present in the 2011 dataset.
Some interesting discrepancies are present in the measured rates of various channels, mainly photons and tau ones, however more statistics will be necessary in order to derive any clear indication on the nature of the resonance.
However, even if the discovery were finally associated with the SM Higgs and the absence of new physics signals were to persist, the interest in extensions of the SM would not fade.
In fact, it will take much more effort to determine if the resonance discovered by the experiments is the SM Higgs or not.
Furthermore, the discovery of new particles in the messy hadronic environment at the LHC experiments can be more challenging than we hoped for: new physics may well require more statistics or more sophisticated search strategies.

A compelling evidence of new physics has been around since 80 years: it is now confirmed in a wide range of observations, from astrophysical observations of the rotation curves in galaxies, through weak gravitational lensing and the Cosmic Microwave Background fluctuations, up to simulations of the galaxy formations in the early Universe, that we need a large amount of Dark Matter in the Universe.
While a new paradigm for our understanding of the dynamics of the Universe may be required, the puzzle can be easily solved by the presence of a stable massive new particle, unaccounted for in the SM, which, besides gravitation, interacts weakly with standard matter.
Motivated by this observation, many models of new physics have been proposed which have a Dark Matter candidate: a rich zoology of such reclusive beasts can be found in the literature.
A common ground is the presence of a discrete symmetry that stabilises the Dark Matter particle and prevents or strongly suppresses decays to SM states: such symmetry is usually added {\it ad hoc}, and in the best cases is accidental or it carries along other benefits for the model.
For instance, in supersymmetry, R-parity can also prevent baryon number violating terms in the supersymmetric Lagrangian, therefore forbidding proton decay.
Extra dimensions have a natural set of symmetries that may play the role of the Dark Matter parity: namely symmetries of the compact space, or Kaluza-Klein (KK) parities~\cite{KKparity}.
However, it has been shown in~\cite{Cacciapaglia:2009pa} that most scenarios studied in the literature require additional conditions on the 4-dimensional counter-terms localised on the boundaries or singular points of the space, therefore the parity is not a direct consequence of the extra dimensional set-up and should be generically considered {\it ad-hoc}.
Furthermore, the KK-parity is easily broken once interesting models are constructed: for instance, Gauge-Higgs unification models require the localisation of fermion zero modes on 4D fixed points~\cite{GHU}, thus potentially breaking the symmetry.

For instance, in 5 dimensional models compactified on an interval $S^1/\mathbb{Z}_2$, the KK parity is given by a translation by the length of the interval (frequently defined as an inversion with respect to the centre of the interval, which is an equivalent definition but which do not act properly on the fields). 
The introduction of a bulk mass term for fermions, which is odd under the $\mathbb{Z}_2$ that defines the orbifold and would localise the chiral zero mode on either end of the interval depending on the sign of the mass term, has been shown to be incompatible with the inversion symmetry~\cite{BarbieriCreminelli}. 
The main issue seems to be related with the presence of fixed points or boundaries, which are however a common feature to all compactifications with chiral zero modes.
This drawback can be turned into a powerful tool by the observation that the requirement of an exact KK-parity is very restrictive and can be used as a powerful selection criterion to single out promising compact spaces.
This criterion has been used in~\cite{Cacciapaglia:2009pa} to show that there exists only one 6D orbifold in 5 or 6 flat dimensions that possesses such a symmetry: the real projective plane (RP$^2$).

Some aspects of the phenomenology of the RP$^2$ have been discussed in~\cite{Cacciapaglia:2009pa} and ~\cite{Cacciapaglia:2011hx}.
In this paper we will focus on the even tiers $(2,0)$ and $(0,2)$: they are important because they lead to events without missing energy, and therefore they can give signatures very different from supersymmetric models.
Furthermore, they may play a crucial role in this model where naturally small mass splittings are generated by loop corrections, thus the phenomenology of the lightest odd tier resembles a compressed spectrum supersymmetric scenario~\cite{compressedSUSY,jetmet}. 
Due to the large interest given to supersymmetric models (both ATLAS and CMS have a dedicated team on supersymmetric, while all other models are binned in the Exotica group), many of these signatures have not been thoroughly explored.
In this paper, we will discuss for the first time the effect of the two different radii allowed by the compactification and the structure of the symmetries of the space.
Then we will introduce the results for the one loop corrections to the masses and discuss in details the spectrum as a function of the two radii.
This discussion is crucial in the understanding of the LHC phenomenology of the model.

The paper is organised as follows: after introducing the model in Section~\ref{sec:intro}, we present a detailed discussion of the mass spectrum of the tiers (2,0) and (0,2) in Section~\ref{sec:spectrum}. In Section~\ref{sec:num} we discuss how we implemented the model in Monte Carlo generators, and in Section~\ref{sec:LHC} we show a series of estimates of the bounds from present LHC searches before concluding in Section~\ref{sec:concl}.

\section{The real projective plane}
\label{sec:intro}

We will consider a quantum field theory defined on a $d$ dimensional flat manifold chosen to be the direct product of the standard four-dimensional Minkowski space-time $\mathcal{M}^4$ and a $d-4$-dimensional orbifold\footnote{An orbifold is defined as a quotient space of a manifold modulo a discrete symmetry group}. 
Our aim is to consider orbifolds without fixed points. Fixed points are not dangerous by themselves, however, as in general they break the $d$-dimensional Lorentz invariance down to the 4-dimensional one, the predictivity of the theory defined on an orbifold with fixed points will be limited. This limitation arises from the fact that the divergences appearing when calculating loop corrections require counter-terms localised on these points. Furthermore, any symmetry associated with the geometry of the space, including the eventual KK-parity, will be broken by generic localised terms unless one imposes some {\it ad hoc} symmetry conditions relating the singularities. As the extra dimensional theories are appealing mainly because they provide a stable dark matter candidate we want the symmetry preserving the stability of the dark matter particle to be an inherent property of the space.

The simplest case would be to study the 1-dimensional orbifolds, that is a circle or an interval. The circle is the simplest orbifold that can be defined as $S^1 = \mathbb{R}/Z$, that is by identifying the points $y \sim y + 2 \pi R$ on an infinite line $\mathbb{R}$ ($R$ is the radius of the circle and $y$ the extra space co-ordinate).
It is known that this orbifold has no fixed points, but no chiral fermion can be defined on it. Therefore this possibility is ruled out.
Chiral fermions can be defined on the interval $S^1/\mathbb{Z}_2$, where the $\mathbb{Z}_2$ symmetry identifies points $y \sim -y = 2 \pi - y$: this orbifold, however, has two fixed points at the boundaries of the interval, $y = 0$ and $y = \pi R$.
From a geometrical point of view, the two fixed points are in-equivalent: they are defects of the space (or branes) where localised 4-dimensional interactions can be added.
Such interactions, which are required as counter-terms for logarithmic divergences at one loop, do violate all the symmetries of the extra dimension, therefore no KK parity will in general survive.
The bulk of the interval is invariant under a mirror symmetry around the centre of the interval, $y \to \pi R - y$, which however interchanges the two fixed points.
Therefore, imposing by hand that the localised interactions on the two fixed points are equal, a KK parity can be obtained: the mirror symmetry is not a good symmetry because the two chiralities of massive fermions would have opposite parity, however it is equivalent to a translation $y \to y + \pi R$ combined with the $\mathbb{Z}_2$ symmetry that defines the orbifold.
This scenario has been studied in the literature~\cite{UED5}, and it was the first example of Dark Matter in extra dimensions: however, the symmetry is imposed {\it ad hoc} and it is easily broken, for instance by bulk mass terms for the fermions which control the localisation of the massless zero modes.

The other face of the coin is that the requirement of spaces having an exact KK parity is a very effective selection rule on the number of viable orbifolds. 
As it was pointed out in~\cite{Cacciapaglia:2009pa} there is a unique 2-dimensional orbifold, among the 17 in-equivalent orbifolds that can be defined on a 2-dimensional euclidean plane, that has no fixed points and such that massless chiral fermionic fields can be defined on it: the real projective plane.
The real projective plane is a compact, non-orientable orbifold of Euler characteristic 1 without boundaries. 
Among all possible descriptions of the real projective plane that are topologically equivalent spaces, we choose the simplest case with flat bulk metric: the spherical projective plane has been considered in~\cite{Dohi:2010vc}, and it leads to a completely different phenomenology.

\subsection{Definition of the ``flat'' RP$^2$ orbifold}

\begin{figure}[tb]
\begin{center}
\includegraphics[scale=0.7]{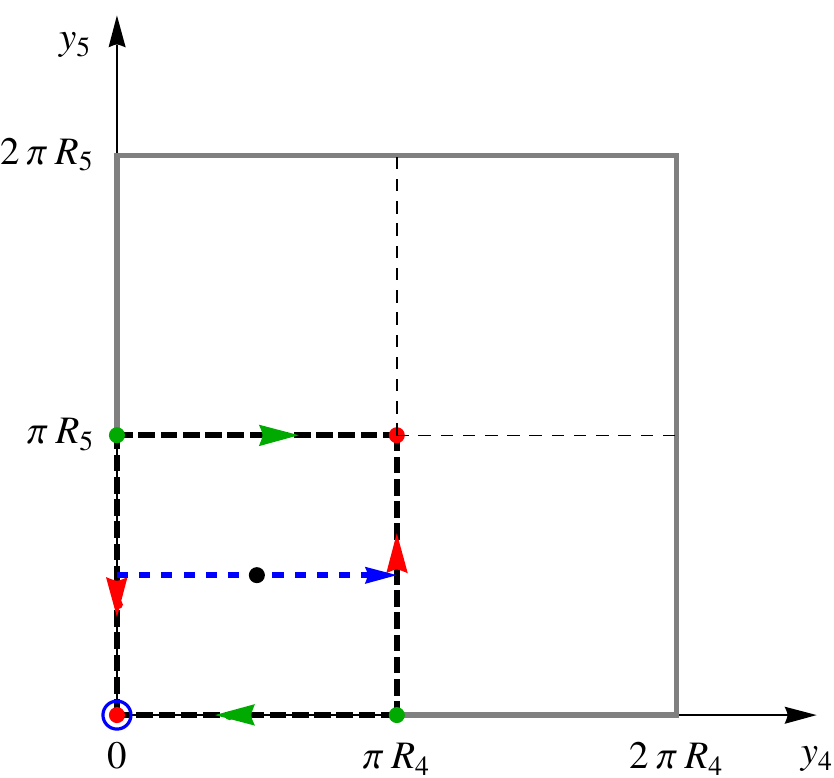}~~~~~~~~~~~~~
\includegraphics[scale=0.7]{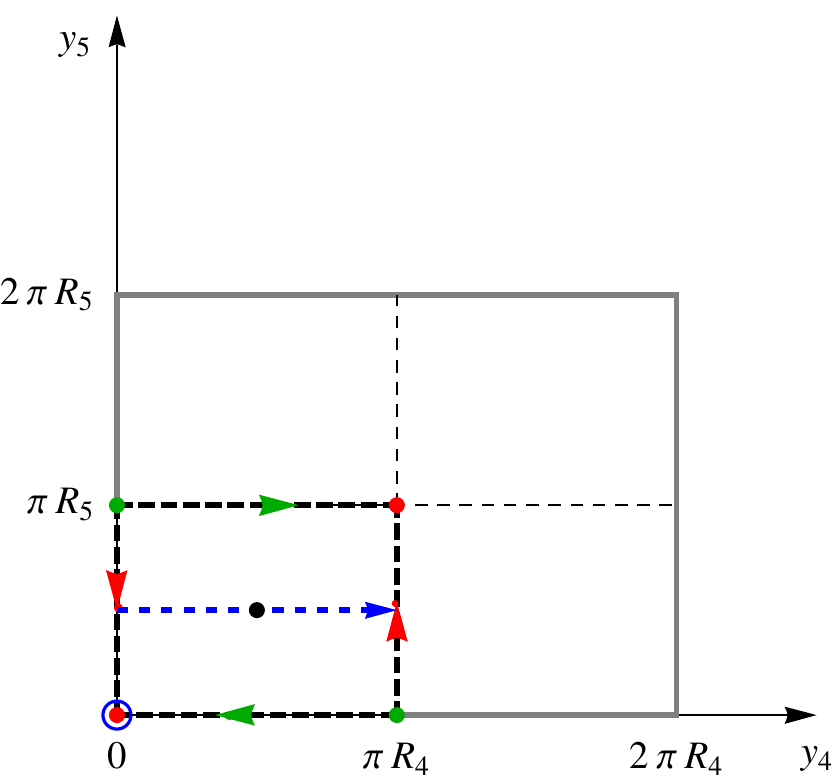}
\end{center}
\caption{\footnotesize The fundamental domain of the flat real projective plane of side lengths of $\pi R_4$ and $\pi R_5$ (within black thick dashed lines) embedded in a torus (grey square). The symmetry generators are represented as blue circle at the origin $(0,0)$ for the $\pi$-rotation generator $r$, and as a blue dashed arrow -  glide generator $g$. Red and green dots represent the identified singular points of the orbifold. The green and red arrows indicate the identification of the edges. The centre of the rectangle (black dot) represents the $P_{KK}$ parity centre. On the left panel we show the symmetric case $\xi=1$ ($R_4 = R_5$), on the right panel the asymmetric orbifold with $\xi=4/3$ ($3 R_4 = 4 R_5$).} \label{fig:RPPdomain}
\end{figure}

The flat real projective plane is described as a quotient space of a plane $\mathbb{R}^2$ modulo a discrete symmetry group\footnote{Note that the structure of the group is entirely defined by the relations between the generators. The particular representation of the generators, in terms of isometries acting on the plane, is not necessary but helps in visualisation.}:
\begin{equation}
 \Gamma_{\rm RP^2} = \langle r,g | r^2 = (g^2*r)^2 = \mathbf{1} \rangle \label{eq:gammaRPP}
\end{equation}
where $r$ is the rotation of $\pi$--degrees around the origin of the co-ordinate system, and $g$ is a glide-reflection:
\begin{equation}
r:\left\{\begin{array}{l}
         y_4 \sim r(y_4) = -y_4 \\
         y_5 \sim r(y_5) = -y_5
        \end{array}\right.\,;  \qquad
g:\left\{\begin{array}{l}
         y_4 \sim g(y_4) = -y_4 +\pi R_4\\
         y_5 \sim g(y_5) =  y_5 + \pi R_5
        \end{array}\right.\,.
\end{equation}
Translations along the two directions are generated by the glide, therefore the space can be seen as a subspace of a 2-dimensional torus with radii $R_4$ and $R_5$.
In this sense this compact space can be thought of as a patch that covers the entire torus once replicated by the symmetries in $\Gamma_{\rm RP^2}$.
Fermions can therefore be defined here in the same way as they are defined on a torus.
One can imagine the fundamental domain of the real projective plane, obtained by applying the above identifications of points of $\mathbb{R}^2$, as a rectangle whose sides are $\pi R_4$ and $\pi R_5$ long, and where facing sides are identified in opposite directions, see Figure~\ref{fig:RPPdomain}.

It is easy to see that this orbifold has no fixed points nor fixed lines. While the rotation $r$ leaves the corners of the fundamental rectangle fixed, opposite corners are identified by the glide: 
$(0,0) \sim (\pi R_4, \pi R_5)$ and $(0,\pi R_5) \sim (\pi R_4,0)$. The two points, that we label $0$ and $\pi$ respectively, are two conical singularities with deficit angle $\pi$, where the metric is still finite.
In general 4-dimensional Lagrangians can be localised on such singular points, and they are required by the divergences of loop corrections: however, the identifications provided by the glide will make sure that some residual symmetries of the extended Poincar\'e group will survive and play the role of KK parity. 

Note that a second glide can be defined as the combination of glide and rotation:
\begin{equation}
g' = g*r:\left\{\begin{array}{l}
         y_4 \sim g(y_4) = y_4 +\pi R_4\\
         y_5 \sim g(y_5) = - y_5 + \pi R_5
        \end{array}\right.
\end{equation}
We can therefore equivalently define the space in terms of the two glides $g$ and $g'$.

The space has two free parameters, i.e. the two radii $R_4$ and $R_5$.
In the following we will use a convenient parametrisation: calling $R_4$ always the largest one of the radii ($R_4 \geq R_5$), we define
\begin{equation}
m_{KK} = \frac{1}{R_4} \equiv \frac{1}{R}\qquad \mbox{and} \quad \xi \equiv \frac{R_4}{R_5} \geq 1\,.
\end{equation}
In this parametrisation, $m_{KK}$ will set the mass scale for the KK resonances, while $\xi$ will parametrise the asymmetry of the rectangle, with $\xi=1$ describing the symmetric case (square).

\subsection{Symmetries of the ``flat'' RP$^2$}

Geometrical symmetries of the fundamental space of the orbifold may be interpreted as parities or global symmetries acting on the Kaluza-Klein modes of the model.
However, this is not always the case, and one needs to be careful with the relation between such symmetries and the symmetries defining the orbifold. In what follows we will keep denoting the generators of the symmetries defining the orbifold by lowercase letters $r$ and $g$ (and $g'$), while we will use capital letters for the symmetries of the fundamental domain.

The fundamental domain of the RP$^2$ is a rectangle with sides $\pi R_4$ and $\pi R_5$.
The corners of the rectangle are special points corresponding to conical singularities of the space where localised counter-terms emerge: in the fundamental space there are only two such points as facing corners are identified by the glide and therefore are identical points.
Also, opposing sides of the rectangle are identified, as shown in Figure~\ref{fig:RPPdomain}.

The fundamental space is invariant under a rotation by $\pi$ degrees around the centre of the rectangle: this symmetry maps the rectangle onto itself, including singular points and sides.
In terms of the co-ordinates:
\begin{equation}
R':\left\{\begin{array}{l}
         y_4 \to R'(y_4) = - y_4 +\pi R_4\\
         y_5 \to R'(y_5) = - y_5 + \pi R_5
        \end{array}\right.\,.
\end{equation}
As this symmetry changes sign to both co-ordinates, the components of the 6-dimensional fermions with opposite chiralities will necessary pick up opposite parity: therefore, this symmetry is not a convenient definition of a parity, because it does not allow do define a single parity to the massive KK fermions.
However, one can define an equivalent symmetry by combining it with the rotation in the orbifold symmetry group:
\begin{equation}
P_{KK} = R'*r:\left\{\begin{array}{l}
         y_4 \to P_{KK}(y_4) = y_4 +\pi R_4\\
         y_5 \to P_{KK}(y_5) = y_5 + \pi R_5
        \end{array}\right.\,.
\end{equation}
Now, the symmetry is a simple translation along both directions, and it can be easily seen that all KK modes in the tier $(k,l)$ will pick up the same phase $(-1)^{k+l}$.
As such, this is a perfect example of KK parity: states with $k+l$ odd will not be able to decay into SM states and the lightest one, belonging to the $(1,0)$ tier will be exactly stable~\footnote{In the case $R_4 = R_5$, there may be two degenerate DM candidates, as the $(0,1)$ is degenerate with $(1,0)$ at three level. More details will be discussed in the following sections.}.
The symmetry does not rely on any assumption on the Lagrangian terms localised on the two singular points, therefore it is also respected by the UV completion of the model.

Imposing conditions on the localised Lagrangians may enhance the symmetries of the model.
For instance we may assume that the Lagrangians localised on the two singular points are the same: this is true, for instance, for the counter-terms generated by loops involving bulk couplings.
In this case, the space is also invariant under a mirror symmetry with respect to the axes $y_4 = \pi R_4/2$:
\begin{equation}
M_4:\left\{\begin{array}{l}
         y_4 \to M_4(y_4) = - y_4 +\pi R_4\\
         y_5 \to M_4(y_5) = y_5
        \end{array}\right.\,.
\end{equation}
Once more, one can remove the minus sign in the transformation of $y_4$ by combining it with the glide $g$:
\begin{equation}
P_{KK}' = M_4*g:\left\{\begin{array}{l}
         y_4 \to P_{KK}'(y_4) = y_4\\
         y_5 \to P_{KK}'(y_5) = y_5 + \pi R_5
        \end{array}\right.\,.
\end{equation}
This is a good parity because all states in the tier $(k,l)$ will pick up the same phase $(-1)^l$.
If this parity were conserved, then transitions like $(0,1) \to (1,0)$ or $(1,1) \to (0,0)$ would be forbidden.
This implies that processes that violate $P'_{KK}$ can not be mediated by bulk loops, and only by (unequal) localised Lagrangians.

The case $R_4 = R_5$ ($\xi = 1$, degenerate RP$^2$) also seems to have enhanced symmetry: the masses of symmetric tiers $(k,l)$ and $(l,k)$ become equal at tree level and the fundamental domain becomes a square.
The square is invariant under a mirror symmetry with respect to the diagonal:
\begin{equation}
M_d:\left\{\begin{array}{l}
         y_4 \to M_d(y_4) = y_5\\
         y_5 \to M_d(y_5) = y_4
        \end{array}\right.\,.
\end{equation}
Such symmetry exchanges symmetric states, therefore the spectrum seems to achieve an enhanced symmetry.
However this is not the case: to see this, we can notice that the exchange of the two directions will also exchange the two glides, $g \leftrightarrow g'$.
Fields which have different parity under the two glides ($p_{g'} = p_g p_r$), therefore, will not be mapped into themselves by this symmetry: this always happens for fields odd under the rotation.
To see this explicitly, let's consider the KK expansion of a scalar field with parities $p_r = -1$ and $p_g = 1$ (therefore, $p_{g'} = - p_g$): the spectrum contains tiers $(2k,0)$ but not $(0,2k)$~\cite{Cacciapaglia:2009pa}, therefore the map cannot be closed. Similarly, tiers $(0,2k-1)$ do not have a symmetric state.
Even if one choses to include only bosons with parity $p_r = +1$ (for which $p_{g'} = p_g$), problems arise when considering fermions.
In fact, the left- and right-handed components of the 6D fermions do have opposite parities under the rotation, therefore it is unavoidable to have fields with $p_r = -1$.
Nevertheless, the spectrum is symmetric and gauge interactions do not depend on the $p_g$ of the field.
On the other hand, Yukawa couplings are proportional to $p_g$, therefore the symmetry $m_d$ will flip the sign of the Yukawa couplings.
There are other localised interactions that may distinguish between the two 6D chiralities and depend on $p_g$.
For such reasons, $M_d$ cannot be considered as an exact symmetry of the orbifold.

To summarise our discussion, we showed that, independently on the value of the radii, the RP$^2$ possesses a unique exact KK parity $P_{KK} = (-1)^{k+l}$.

In the following, we will consider the space-time $\mathcal{M}^4 \times$ RP$^2$ with a 6-dimensional flat metric $g^{MN} = diag(1,-1,-1,-1,-1,-1)$. 
We denote by $x^\mu$ the co-ordinates of $\mathcal{M}^4$ with $\mu \in \{0,1,2,3\} $ and by $y^\alpha$ the co-ordinates on the orbifold with
$\alpha \in \{4,5\}$. When needed we note a vector $\vec{y} = (y_4, y_5)$. 
The capital letters of the alphabet run over all the co-ordinates of the six-dimensional space, $M,N \in \{0,1,2,3,4,5 \}$.

\subsection{Parities of the fields and KK decomposition}

In this section we summarise the Kaluza-Klein decomposition of quantum fields defined on the six-dimensional background 
$\mathcal{M}^4 \times RP^2$ described in the previous section. We introduce the fields defined in six dimensions: 
a scalar field $\Phi(x^\mu, y^\alpha)$, a gauge field $A_M(x^\mu, y^\alpha)$ and a spinor $\Psi(x^\mu, y^\alpha)$. 
For every orbifold symmetry, namely the rotation $r$ and the glide $g$ in our case,  it corresponds a transformation of the fields:
\begin{equation}
X(r(y^\alpha)) = p_r \mathcal{R} X(y^\alpha)\, \qquad \mbox{and} \quad X(g(y^\alpha)) \rightarrow p_g \mathcal{G} X(y^\alpha)\,,
\end{equation}
where $\mathcal{R}$ and $\mathcal{G}$ are the transformation operator in the adequate spin representation and $X$ stands for any quantum field.
The invariance under the orbifold projection requires to impose such properties on each field in the theory, where the two parities $(p_r, p_g)$ determine the spectrum and property of the field. 
From the properties of the symmetries $r$ and $g$ in Eq.~\ref{eq:gammaRPP}, it can easily be shown that the possible parities of the fields under the rotation and glide-reflection are $(p_r,p_g) = (\pm, \pm)$ respectively~\footnote{The glide can also pick up an imaginary phase, thus $(p_r,p_g) = (\pm, \pm i)$ are also allowed choices. We will not consider them here for simplicity.}. 
Thus every field will be characterised by a pair $(p_r,p_g)$. 
As the RP$^2$ has no boundaries, we can decompose each field in KK modes as on the torus, where the wave functions are combinations of sine and cosine functions of the extra co-ordinates, and then impose the orbifold projection on each field.

\subsubsection*{Scalar field}
As a simple example, we can detail the case of a scalar field, for which $\mathcal{R}$ and $\mathcal{G}$ are equal to the identity operator.
The 6D action of a complex scalar field $\Phi(x^\mu, y^\alpha)$ of mass $M_\Phi$ is~\footnote{In this action we integrate over the fundamental domain of the torus. Integrating over the domain of the RP$^{2}$ would only change the normalisation of the fields, however the physical properties, and the KK expanded Lagrangians, are exactly the same.}
\begin{equation}\label{Sscalar}
 S_{\Phi}^{6D} = \int d^4x^\mu \int_{0}^{2\pi R_4} dy^4 \int_{0}^{2\pi R_5} dy^5 \left[ 
D_M \Phi^\dagger D^M \Phi - M_\Phi^2 \Phi^\dagger\Phi \right]
\end{equation}
with $D_M = \partial_M -i g A_M^a t_r^a$ the covariant derivative. 
The field is decomposed into KK states labelled by a pair of positive numbers $(k,l)$, with $k,l \geq 0$
\begin{equation}
 \Phi(x^\mu, y^\alpha) = \sum_{k,l \geq 0} \phi^{(k,l)}(x^\mu)f_{k,l}(y^\alpha)\,,
\end{equation}
where the wave function is generically a combination of $\sin$ and $\cos$: 
\begin{equation}
f_{k,l}(y^\alpha) \propto [\sin(k y_4/R_4) \,, \cos(k y_4/R_4)] \times  [\sin(l y_4/R_4) \,, \cos(l y_4/R_4)]\,.
\end{equation}
Solving the equations of motion, one finds the mass spectrum of the Kaluza-Klein resonances at tree level :
\begin{equation}
 \left(-\partial_4^2- \partial_5^2 + M_\Phi^2\right) f_{k,l}(y^\alpha) = M_{k,l}^2 f_{k,l}(y^\alpha) = \left( m_{k,l}^2 + M_\Phi^2\right) f_{k,l}(y^\alpha)
\end{equation}
with
\begin{equation}
 m_{k,l}^2 = \frac{k^2}{R_4^2} + \frac{l^2}{R_5^2} = \sqrt{k^2 + \xi^2 l^2}\; m_{KK}\,.
\end{equation}

The functions $f_{k,l}$ are found by imposing the parities $(p_r,p_g)$ on the wave functions, so there are four possible KK towers 
with different parities:
\begin{eqnarray}
\Phi^{(++)} & = & \frac{1}{2 N}\phi^{(0,0)} + \frac{1}{\sqrt{2} N}\sum_{k=1}^\infty\left(\cos(2ky_4/R_4)~\phi^{(2k,0)} + \cos(2ky_5/R_5)~\phi^{(0,2k)}\right) +\\
 & + & \frac{1}{N}\sum_{k,l > 0}\left( \cos(ky_4/R_4)~ \cos(ly_5/R_5)~ \phi^{(k,l)}_{k+l=2m} + \sin(ky_4/R_4) ~\sin(ly_5/R_5)~ \phi^{(k,l)}_{k+l=2m+1}\right)\,, \nonumber\\
\Phi^{(+-)} & = &\frac{1}{\sqrt{2} N}\sum_{k=1}^\infty\left(\cos((2k-1)y_4/R_4)~\phi^{(2k-1,0)} + \cos((2k-1)y_5/R_5)~\phi^{(0,2k-1)}\right) + \\
 & + & \frac{1}{N}\sum_{k,l > 0}\left( \sin(ky_4/R_4)~ \sin(ly_5/R_5)~ \phi^{(k,l)}_{k+l=2m} + \cos(ky_4/R_4) ~\cos(ly_5/R_5)~ \phi^{(k,l)}_{k+l=2m+1}\right)\,, \nonumber\\
\Phi^{(-+)} & = &\frac{1}{\sqrt{2}N}\sum_{k=1}^\infty\left(\sin(2ky_4/R_4)~\phi^{(2k,0)} + \sin((2k-1)y_5/R_5)~\phi^{(0,2k-1)}\right)  +\\
 & + & \frac{1}{N}\sum_{k,l > 0}\left( \sin(ky_4/R_4)~ \cos(ly_5/R_5)~ \phi^{(k,l)}_{k+l=2m} + \cos(ky_4/R_4) ~\sin(ly_5/R_5)~ \phi^{(k,l)}_{k+l=2m+1}\right)\,, \nonumber\\
 \Phi^{(--)} & = &\frac{1}{\sqrt{2}N}\sum_{k=1}^\infty\left(\sin((2k-1)y_4/R_4)~\phi^{(2k-1,0)} + \sin(2ky_5/R_5)~\phi^{(0,2k-1)}\right)  +\\
 & + & \frac{1}{N}\sum_{k,l > 0}\left( \cos(ky_4/R_4)~ \sin(ly_5/R_5)~ \phi^{(k,l)}_{k+l=2m} + \sin(ky_4/R_4) ~\cos(ly_5/R_5)~ \phi^{(k,l)}_{k+l=2m+1}\right)\,; \nonumber
\end{eqnarray}
where $N = \pi \sqrt{R_4 R_5}$.
The only case with a zero mode is $(+,+)$, therefore a model with bulk Higgs must have such a parity in order to include a Higgs boson in the zero mode spectrum.

\subsubsection*{Gauge fields}
Gauge fields in 6 dimensions are given by a 6-component vector:
$$A_M(x^\mu,y^\alpha) = ( A_\mu(x^\mu,y^\alpha), A_\beta(x^\mu,y^\alpha) )\,.$$ 
Being a vector, the parities of the various component of $A_M(x^\mu,y^\alpha)$ are sensitive to the action of the symmetry on the co-ordinates.
In fact, under rotation and glide, a 6-vector transforms as
$$
A_M(r(y^\alpha)) = \mathcal{R} ( A_\mu(y^\alpha), - A_4 (y^\alpha), - A_5 (y^\alpha) )\,, \quad A_M(g(y^\alpha)) = \mathcal{G} ( A_\mu(y^\alpha), - A_4 (y^\alpha), A_5 (y^\alpha) )\,,
$$
where $\mathcal{R}$ and $\mathcal{G}$ are suitable gauge transformations.
In the following we will focus on a model where the gauge groups are the same as in the SM, therefore we will require that all the gauge fields have a vector zero mode, associated with an unbroken 4D gauge symmetry.
Therefore, for all the gauge groups, it is enough to chose $\mathcal{R}$ and $\mathcal{G}$ equal to the identity.
More in general, a non trivial gauge transformation will allow to break the 6D gauge symmetry to a subgroup in 4D (namely, not all the gauge bosons will have a massless vector mode, but only the gauge fields associated with a subgroup).

For a $(+,+)$ vector, which will contain a zero mode vector boson in the KK theory, the parity of the ``scalar'' components will be $(-,-)$ for $A_4$ and $(-,+)$ for $A_5$.
The spectrum of gauge fields is complicated by the mixing between the vector components and the scalar ones, and an arbitrary gauge choice. A complete and detailed analysis of the spectrum can be found in~\cite{Cacciapaglia:2009pa}. Here we will limit ourselves to a brief description of the spectrum in the Feynman--'t Hooft gauge, where the goldstone bosons appear explicitly in the spectrum with the same mass as the vectors they belong to. The spectrum of the $A_\mu$, $A_4$ and $A_5$ components is the same as for a scalar field listed before.
The massive vectors in the spectrum pick up their longitudinal degree of freedom by eating up a linear combination of the two scalar components.
While the zero mode is left massless, the $(2k,0)$ vector modes eat up the $(2k,0)$ modes in $A_5$ while the $(0,2k)$ ones eat up the $(0,2k)$ modes in $A_4$.
On the other hand, the $(2k-1,0)$ modes in $A_4$ and the $(0,2k-1)$ modes in $A_5$ are physical scalars because there is no corresponding vector boson with the same mass.
For the $(k,l)$ modes, a linear combination of $A_4$ and $A_5$ is eaten, while another one is left in the spectrum.

\subsubsection*{Spinor field}
The spectrum for fermions is also complicated by the action of the orbifold symmetries on the components of the fermion field, and we invite the reader to consult Ref.~\cite{Cacciapaglia:2009pa} for a detailed description. Here we mention only some points that are important for our purposes. In 6 dimensions  it is possible to define two independent 6D chirality operators (which are different from the 4-dimensional chirality!). A generic 6D fermion has four components that are Weyl spinors: $\Psi^{6D} = (\chi_+,\,\bar{\eta}_-,\,\chi_-,\,\bar{\eta}_+)^T$ and in this notation $\pm$ subscripts correspond to the 6D chiralites while $\chi$, $\eta$ are the 4D chirality eigenstates. The glide symmetry flips the 6D chirality of the fermion it is acting on as

\begin{equation}
p_g \mathcal{G} \Psi(y^\alpha) = p_g \mathcal{G} (\chi_+,\,\bar{\eta}_-,\,\chi_-,\,\bar{\eta}_+)^T = p_g (\chi_-,\,\bar{\eta}_+,\,\chi_+,\,\bar{\eta}_-)^T\,,
\end{equation}
therefore a consistent model with fermions must contain 6D vector-like fermions.
The parity under the glide $p_g$ does not play any significant role and it's only constrained by the presence of Yukawa couplings because the $p_g$ of the Higgs field is fixed to be equal to $+1$.
On the other hand, a bulk mass, that would connect the two 6D chiralities, is forbidden by the rotation symmetry.
For a vector-like fermion in 6D, the parity under rotation selects the chirality of the zero mode~\footnote{Note that the phase under rotation of the left-handed and right-handed components of the 6D fields are opposite: $p_r \mathcal{R} \Psi(y^\alpha) = p_r \mathcal{R} (\chi_+,\,\bar{\eta}_-,\,\chi_-,\,\bar{\eta}_+)^T = p_r (\chi_+,\,-\bar{\eta}_-,\,\chi_-,\,-\bar{\eta}_+)^T$. What we define as the parity of the 6D field is the parity of its left-handed components.}: $p_r = +1$ for a left-handed zero mode and $p_r = -1$ for a right-handed one.
For each choice of the parities, the KK tower of a fermion will contain a chiral zero mode, one vector-like fermion in each tier $(k,0)$ and $(0,k)$ and two vector-like fermions in each $(k,l)$ tier. 
Note also that for each SM fermion, the 6D theory will contain an independent vector-like 6D fermion for the two chiralities: for instance, the electron will be associates with a 6-dimensional spinor $e_L^{6D}$ with $p_r= +1$ corresponding to the left handed component of the electron, and a $e_R^{6D}$ with $p_r= -1$ corresponding to the right handed component.

\begin{table}[tb]
\begin{center}
\begin{tabular}{|l||*{6}{c|}}
\hline
 & SU(3)$_c\times$SU(2)$_w\times$U(1)$_Y$ & $(p_r, p_g)$ & $(0,0)$ & $\begin{array}{c} (2k-1,0) \\ (0,2k-1) \end{array}$ & $\begin{array}{c} (2k,0) \\ (0,2k) \end{array}$ & $(k,l)$ \\
\hline
\hline
$G_\mu$ &  \multirow{2}{*}{$(\mathbf{8}, \mathbf{1}, 0)$}  & \multirow{2}{*}{$(+,+)$} & $\surd$ & & $\surd$ & $\surd$ \\
$G_\phi$ &  & & & $\surd$ & & $\surd$ \\
\hline
$W_\mu$  &  \multirow{2}{*}{$(\mathbf{1}, \mathbf{3}, 0)$} & \multirow{2}{*}{$(+,+)$} & $\surd$ & & $\surd$ & $\surd$ \\
$W_\phi$&  & & & $\surd$ & & $\surd$ \\
\hline
$B_\mu$ &  \multirow{2}{*}{$(\mathbf{1}, \mathbf{1}, 0)$} & \multirow{2}{*}{$(+,+)$} & $\surd$ & & $\surd$ & $\surd$ \\ 
$B_\phi$&  & & & $\surd$ & & $\surd$ \\
\hline
$L$ & $(\mathbf{1}, \mathbf{2}, -1/2)$ & $(+,p_g^l)$ & $\surd$ l.h. & $\surd$ & $\surd$ & $\surd$$\surd$ \\ 
\hline
$E$ & $(\mathbf{1}, \mathbf{1}, -1)$ & $(-, p_g^l)$& $\surd$ r.h. & $\surd$ & $\surd$ & $\surd$$\surd$ \\ 
\hline
$Q$ & $(\mathbf{3}, \mathbf{2}, 1/6)$ & $(+, p_g^q)$& $\surd$ l.h. & $\surd$ & $\surd$ & $\surd$$\surd$ \\ 
\hline
$U$ & $(\mathbf{3}, \mathbf{1}, 2/3)$ & $(-,p_g^q)$& $\surd$ r.h. & $\surd$ & $\surd$ & $\surd$$\surd$ \\ 
\hline
$D$ & $(\mathbf{3}, \mathbf{1}, -1/3)$ & $(-,p_g^q)$& $\surd$ r.h. & $\surd$ & $\surd$ & $\surd$$\surd$ \\ 
\hline
$H$ & $(\mathbf{1}, \mathbf{2}, 1/2)$ & $(+,+)$& $\surd$ & & $\surd$ & $\surd$ \\ 
\hline
\end{tabular}
\caption{\footnotesize UED spectrum on the RP$^2$. The subscript $\phi$ indicate the physical gauge-scalar. The parity under glide of leptons $p_g^l$ and quarks $p_g^q$ are arbitrary, and due to the fact that the fermions are 6D vector-like, either choice will lead to the same physics.} \label{tab:UED}
\end{center}
\end{table}

\subsubsection*{UED model}

\begin{figure}[tb]
\begin{center}
\includegraphics[width=11cm]{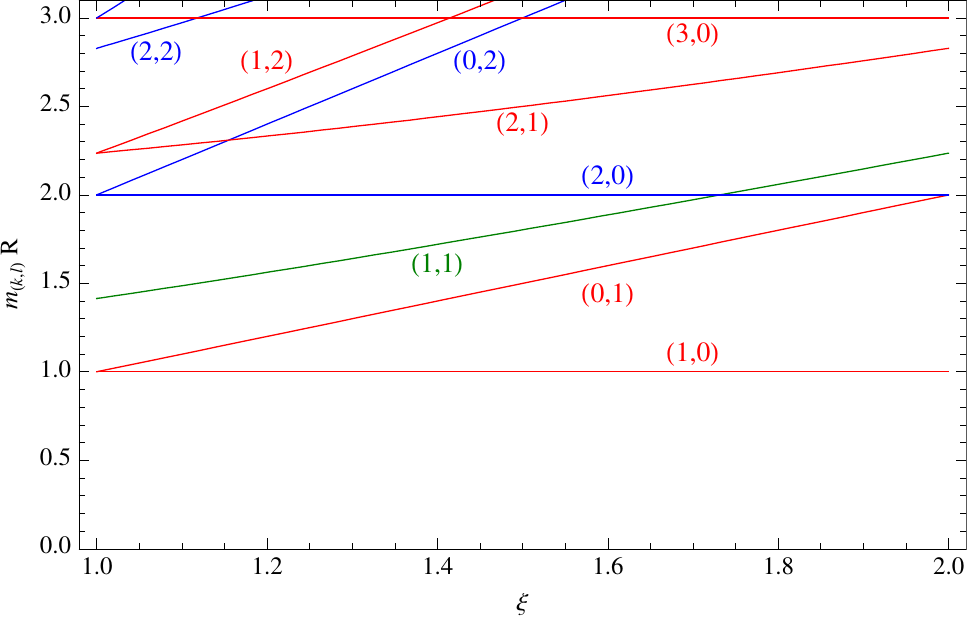}
\end{center}
\caption{\footnotesize Crossover between tiers as a function of $\xi$. The colour-code indicates the KK parity of the tiers: red for odd tiers, blue and green for even ones where the green ones can only decay into zero modes via the localised interactions.} \label{fig:crossover}
\end{figure}

In the following, we will focus on an incarnation of Universal Extra Dimension (UED) models on the RP$^2$: in this model, each SM field is the zero mode of a 6-dimensional field with suitable parities.
The tree-level spectrum of the model is summarised in Table~\ref{tab:UED}.
In the degenerate case $\xi=1$, when $R_4 = R_5 = R$, the symmetric levels $(k,l)$ and $(l,k)$ are exactly degenerate at tree level.
When $\xi > 1$, the degeneracy is removed; moreover, as $\xi$ increases, there are crossovers between different levels.
This fact can be very important for the phenomenology of the model both at the LHC and regarding the Dark Matter abundance calculations.
As shown in Figure~\ref{fig:crossover}, the crossover can take place for small values of $\xi$ among more massive levels.
On the other hand, the crossover for light modes requires $\xi-1$ to be order 1: for instance, the level $(0,1)$ becomes degenerate with $(2,0)$ for $\xi = 2$ and level $(1,1)$ with $(2,0)$ for $\xi = \sqrt{3} \sim 1.7$.

\section{Loop corrections to the level (2,0) and (0,2) masses}
\label{sec:spectrum}

Loop corrections play a crucial role for the phenomenology of UED models, because the states in each tier are exactly degenerate at tree level (ignoring effects from the electroweak symmetry breaking).
Furthermore, for even tiers, loops can also induce vertices that mediate decays directly into a pair of SM particles, thus violating the KK momentum conservation rule.
For the lightest odd tier, the mass splitting will determine the typical visible energy of the decay chain, which will inevitably end into the Dark Matter candidate; for even tiers, the loop corrections are crucial to determine the branching ratios into resonant SM pairs, the relevance of Drell-Yan resonant production via a heavy state, which can impact both the LHC search strategy and add resonant contributions to the Dark Matter annihilation cross sections.
The latter determines the relic abundance of Dark Matter and therefore the preferred mass range to explain current cosmological observations~\cite{RP2DM}.

In the rest of this section, we will introduce the loop corrections to the masses of the even states (2,0) and (0,2), focusing on the dependence on the two radii ($R_4$ and $R_5$, or equivalently $m_{KK}$ and $\xi$) and the relevance of the mixing between the two tiers.
We will also recall the analogous formulas for the odd states (1,0) and (0,1)~\cite{Cacciapaglia:2009pa}, including the dependence on the two radii.
Loop induced KK-number violating vertices and mixing terms between the two even tiers, which also violate KK-number conservation, only depend on the localised divergences, thus they can be extracted by using the localised counter-terms as in~\cite{Cacciapaglia:2011hx}.
In the following, we will only calculate the mass mixing in detail.

\subsubsection*{Mixed propagator formalism}

The loop corrections to modes with a zero KK number, like $(n,0)$ and $(0,n)$ can be conveniently computed by use of mixed propagators in 6 dimensions: we will Fourier-transform the non-compact co-ordinates $x^\mu$ into momentum space $p^\mu$, which contains the physical momentum and energy measured by 4-dimensional observers, and leave the other two co-ordinates in position space.
The 6D propagator can in general be expressed in terms of Bessel functions~\cite{DaRold:2003yi}, which are not very handy in loop calculations.
Using the presence of a zero KK number for the external lines of the loops, we can expand in KK modes along the direction corresponding to the zero external KK momentum.
For instance, for modes $(n,0)$, we can expand in KK modes along the second co-ordinate $y^5$.
The 6D propagators can therefore be written in terms of 5D propagators: for instance, for a scalar propagating on a torus, we have
\begin{equation}
 G_\Phi^{\rm torus}\left( p^\mu, \vec{y}-\vec{y'}\right) = \sum_{l=-\infty}^{\infty} G_\Phi^{\rm circle}\left( \chi_l,|y_4-y_4'|\right) f_l^*(y_5)f_l(y'_5)\,.
 \end{equation}
In this expression, $f_l(y)$ are the normalised wave-functions on the circle and $l$ labels the masses of KK states expanded along $y_5$,
 \begin{equation}
 f_l(y_5) = \frac{ e^{il \frac{y_5}{R_5}}}{\sqrt{2\pi R_5}}\,,
 \end{equation}
and $G_\Phi^{\rm circle}$ are the 5D propagators~\cite{Puchwein:2003jq} of a scalar field of mass $\frac{l}{R_5}$,
 \begin{equation}
 G_\Phi^{\rm circle}\left( \chi_l,|y_4-y'_4|\right) = i \frac{\cos\chi_l(\pi R_4-|y_4-y'_4|)}{2\chi_l \sin (\chi_l \pi R_4)}\,, \quad \mbox{with} \quad \chi_l = \sqrt{p^2 - \frac{l^2}{R_5^2}}\,.
 \end{equation}
The advantage of this expressions is that the integrals on the co-ordinate $y^5$ are very simple as the wave functions of the external states do not depend on it, thus we are left with integrals of exponential functions.
The propagators of vectors and fermions can be written in terms of the scalar propagator: in Feynman-'t Hooft gauge
 \begin{equation}
G^{\rm torus}_{A_M} = - g^{M N} G_\Phi^{\rm torus}\,, \quad \mbox{and} \quad G^{\rm torus}_{\Psi} = (p^\mu \Gamma_\mu - i \Gamma_4 \partial_{y_4} - i \Gamma_5 \partial_{y_5})\,  G_\Phi^{\rm torus}\,,
 \end{equation}
where $\Gamma_M$ are the Dirac Gamma matrices in 6D.

Finally, the propagator on the orbifold can be written in terms of the torus propagator observing that the propagators from $\vec{y}$ to the images of $\vec{y'}$ via the symmetries of the orbifold are physically identical.
Therefore, one obtains~\cite{Georgi:2000ks}:
\begin{eqnarray}\label{RPPpropagator}
  G_\Phi^{RP^2}\left( p^\mu,\vec{y}, \vec{y'}\right) & = & \frac{1}{4}\left[ G_\Phi^{\rm torus}\left( p^\mu,\vec{y}-\vec{y'}\right) + p_g\, G_\Phi^{\rm torus}\left( p^\mu,\vec{y}- g(\vec{y'})\right) \right. \nonumber\\
 & + & \left.p_g p_r\, G_\Phi^{\rm torus}\left( p^\mu,\vec{y}-[g*r](\vec{y'})\right) + p_r\, G_\Phi^{\rm torus}\left( p^\mu,\vec{y}-r(\vec{y'})\right) \right]\,,
\end{eqnarray}
where $(p_g, p_r)$ are the parities of the propagating 6D scalar field.
The factor of $1/4$ takes into account the fact that we are integrating over the fundamental space of the torus, which is 4 times bigger than that of the RP$^2$.
Note that all the wave functions are also normalised over the torus, without affecting the physical results.
For vectors and fermions, we need to additionally multiply by the transformation operators in the appropriate spin representation, thus, for instance
\begin{multline}
  G_{A_M}^{RP^2}\left( p^\mu,\vec{y}, \vec{y'}\right)  =  - \frac{g^{M N}}{4}\left[ G_\Phi^{\rm torus}\left( p^\mu,\vec{y}-\vec{y'}\right) + (-1)^{\delta_{(M,5)}} p_g\, G_\Phi^{\rm torus}\left( p^\mu,\vec{y}- g(\vec{y'})\right) \right. \\
  +  \left. (-1)^{\delta_{(M,4)}}\, p_g p_r\, G_\Phi^{\rm torus}\left( p^\mu,\vec{y}-g*r(\vec{y'})\right) + (-1)^{\delta_{(M,4)}+\delta_{(M,5)}}p_r\, G_\Phi^{\rm torus}\left( p^\mu,\vec{y}-r(\vec{y'})\right) \right]\,. 
\end{multline}

\subsubsection*{General loop structure}

Using this form of propagator as in Eq.(\ref{RPPpropagator}), a generic loop correction to any field can be decomposed as
\begin{equation}\label{GenericLoop}
 i\Pi (m_{KK}, \xi) = \frac{i}{4} \big(\Pi_T (m_{KK}, \xi) + p_g\, \Pi_G (m_{KK}, \xi) + p_g p_r\, \Pi_{GR} (m_{KK}, \xi) + p_r\, \Pi_R (m_{KK}, \xi) \big)\,,
\end{equation}
where the parities refer to any one field propagating in the loop.

The first term in Eq.(\ref{GenericLoop}), $\Pi_T$, corresponds to the loop corrections of a theory defined on a torus: this contribution is generically divergent, however it gives a finite contribution after renormalisation of the bulk kinetic terms. 
Here we will follow the prescription of Ref.s~\cite{Cheng:2002iz,Ponton:2005kx}, where we remove the contribution of the zero winding modes, i.e. the modes that do not wrap around the torus.
The terms $\Pi_G$ and $\Pi_{GR}$ correspond to glide symmetries and are finite. 
The reason behind the finiteness is that the points $\vec{y}$ and its image $g(\vec{y})$  (or $[g*r](\vec{y})$) never coincide, i.e. the glide(s) do not admit any fixed points.
The last term, $\Pi_R$, corresponds to the rotation transformation, and it is divergent because the rotation admits fixed points. It was shown explicitly in \cite{Cacciapaglia:2009pa} that divergences only arise in the points where $\vec{y_\ast} = r(\vec{y_\ast})$, i.e. on the corners of the rectangle.
Thus, such divergences can be renormalised by adding counter-terms on the two singular points, whose structure has been studied in~\cite{Cacciapaglia:2011hx}.
The bulk loops simply require equal terms on the singular points, as loop interactions cannot distinguish between the two.

%
%

The loop contributions generally depends on both radii, i.e. $m_{KK}$ and $\xi$ as in the Eq.\ref {GenericLoop}. 
One can easily relate the loop contributions to the modes $(n,0)$ and $(0,n)$ by simply exchanging the two radii, and the two glides:
\begin{eqnarray} \label{GenericLoop2}
 i\Pi^{(0,n)} (m_{KK}, \xi) &=& \frac{i}{4} \left(\Pi^{(n,0)}_T (\xi m_{KK}, 1/\xi) + p_g\, \Pi^{(n,0)}_{GR} (\xi m_{KK}, 1/\xi) + \right. \nonumber \\
 & & \left. p_g p_r\, \Pi^{(n,0)}_{G} (\xi m_{KK}, 1/\xi) + p_r\, \Pi^{(n,0)}_R (\xi m_{KK}, 1/\xi) \right)\,.
\end{eqnarray}
From Eq.~\ref{GenericLoop2} it is clear that, in the degenerate case $\xi=1$, we have $\Pi^{(n,0)} = \Pi^{(0,n)}$ only if $p_g = p_{g'} = p_g p_r$ (i.e. $p_r = 1$) for all the fields running in the loop because, as we will see, $\Pi_G \neq \Pi_{GR}$. 

The loops will also generate mass mixing between states with different KK numbers, as long as they are allowed by the symmetries of the space.
For instance, states $(n_1,0)$ and $(n_2,0)$ can mix via a one loop diagram of modes $\left( \frac{n_1+n_2}{2},m\right)$ and $\left( \frac{|n_1-n_2|}{2},m\right)$ only if both $n_1$ and $n_2$ are even or odd.
Analogously, modes $(n_1,0)$ and $(0,n_2)$ can mix only for even $n_1$ and $n_2$ (which are allowed to mix with $n=0$).
The mixing effect is negligibly small, unless the modes that mix are degenerate at tree level: in the following we will be particularly interested in the mixing between $(2,0)$ and $(0,2)$ which are degenerate in the equal radii $\xi=1$ case.
The odd modes, like $(1,0)$ and $(0,1)$ cannot mix via loops, however they can mix via asymmetric localised terms.
From Eq. (\ref{GenericLoop}), we can infer that the off-diagonal mixing terms can only be generated by the rotation-term $\Pi_R$, because both the torus and the glides preserve KK number~\footnote{In fact, the orbifold defined by the glide alone is the Klein bottle, whose loop contribution is given by $\Pi_{\rm Klein} = \frac{1}{2} \left( \Pi_T + p_g \Pi_G \right)$. The Klein bottle is invariant under translations along the two directions and has no fixed or singular points, therefore no KK-number violating mixing can be generated by either $\Pi_T$ or $\Pi_G$ (and $\Pi_{GR}$).}.

\subsection{Gauge bosons}

In this section we will present the results for the mass corrections to the tiers $(n,0)$, including novel results for even $n$ (vectors) and summarising the results for $n$ odd (scalar) from~\cite{Cacciapaglia:2009pa}.
The results for $(0,n)$ modes can be extracted by use of Eq.~\ref{GenericLoop2}.
The loop corrections to the masses can be written in general as:
\begin{equation}
\delta m^2_{(n,0)} = \frac{g^2 C(r)}{64 \pi^4} m_{KK}^2 \; A_{(n,0)}\,,
\end{equation}
where $g$ is the gauge coupling, $C(r)$ the Casimir of the representation $r$ of the field running in the loop (to be substituted by the charge squared for a U(1) gauge boson), and the loop coefficients $A_{(n,0)}$ are given in the following table:
\begin{center}
\begin{tabular}{ll|c|c|c|c|}
\multicolumn{2}{c|}{$A_{(n,0)}$} & torus & glide ($p_g \times$) & glide' ($p_g p_r \times$) & rotation ($p_r\times$) \\
 \hline
\multirow{2}{*}{gauge loops} &  $n$-even & \multirow{2}{*}{$4 T_6 (\xi)$} &  \multirow{2}{*}{$14 \zeta(3)$} & $\xi^2 \left( 14 \zeta(3) + V_g (n/\xi) \right)$ & \multirow{2}{*}{$8 n^2 \pi^2 L$}\\
 & $n$-odd & &  & $\xi^2 \left( 14 \zeta(3) + S_g (n/\xi) \right)$ & \\
 \hline
\multirow{2}{*}{fermion loops} &  $n$-even & \multirow{2}{*}{$-8 T_6 (\xi)$} & \multirow{2}{*}{$0$} & \multirow{2}{*}{$0$} & \multirow{2}{*}{$0$}\\
& $n$-odd & &  & & \\
 \hline
\multirow{2}{*}{scalar loops} &  $n$-even & \multirow{2}{*}{$T_6 (\xi)$} & \multirow{2}{*}{$7 \zeta(3)$} & $\xi^2 \left( 7 \zeta(3) + V_s (n/\xi) \right)$ & $-\frac{1}{3} n^2 \pi^2 L$\\
 & $n$-odd & &  &$\xi^2 \left( 7 \zeta(3) + S_s (n/\xi) \right)$ & $n^2 \pi^2 L$\\
 \hline
 \end{tabular} \end{center} 
where $\zeta(3)$ is the Riemann Zeta function, 
\begin{equation}
T_6 (\xi) = \frac{1}{\pi \xi} \sum_{ \tiny \begin{array}{c}
n, m \subset Z \\
\{n, m\} \neq \{0,0\} \end{array}}^{\infty} \frac{1}{(n^2+m^2/\xi^2)^2}
\end{equation}
encodes the contribution of the torus loops (numerically, $T_6 (1) = 1.92\dots$), $L = \log (\Lambda^2/m_{KK}^2)$ encodes the logarithmic dependence on the cut--off $\Lambda$, and the functions $S_{g,s}$ and $V_{g,s}$ contain small and finite corrections due to the glides.
The latter functions and $T_6(\xi)$ can be found in the Appendix~\ref{app:functions}.
From the table it can be seen that the difference between the scalar  ($n$-odd) and vector ($n$-even) components of the gauge fields comes from the contribution of small finite glide terms, and of the logarithmic divergent contribution of scalar loops.

In the UED model we are interested in here, the only bulk fields are the SM ones, with parities $(+,+)$ (except for fermions, whose parities do not enter the mass corrections), therefore one can calculate the mass corrections for the U(1), SU(2) and SU(3) gauge bosons as follows:
\begin{eqnarray}
\delta m^2_{G_\phi} &=& \frac{g_3^2}{64 \pi^4} m_{KK}^2 \left[ - 36 T_6 (\xi) + (1+\xi^2) 42 \zeta(3) + 3 \xi^2 S_g (n/\xi) + 24 n^2 \pi^2 L\right]\, \nonumber \\
\delta m^2_{B_\phi} &=& \frac{g_1^2}{64 \pi^4} m_{KK}^2 \left[ - 79 T_6 (\xi) + (1+\xi^2) 7 \zeta(3) + \xi^2 S_s (n/\xi) + n^2 \pi^2 L\right]\,, \\
\delta m^2_{W_\phi} &=& \frac{g_2^2}{64 \pi^4} m_{KK}^2 \left[ - 39 T_6 (\xi) + (1+\xi^2) 35 \zeta(3) + \xi^2 (2 S_g (n/\xi) + S_s (n/\xi)) + 17 n^2 \pi^2 L\right]\,, \nonumber \label{gauge_corr1}
\end{eqnarray}
for the scalars $n$-odd; and
\begin{eqnarray}
\delta m^2_{G_\mu} &=& \frac{g_3^2}{64 \pi^4} m_{KK}^2 \left[ - 36 T_6 (\xi) + (1+\xi^2) 42 \zeta(3) + 3 \xi^2 V_g (n/\xi) + 24 n^2 \pi^2 L\right]\,
\nonumber \\
\delta m^2_{B_\mu} &=& \frac{g_1^2}{64 \pi^4} m_{KK}^2 \left[ - 79 T_6 (\xi) + (1+\xi^2) 7 \zeta(3) + \xi^2 V_s (n/\xi) - \frac{1}{3} n^2 \pi^2 L\right]\,, \\
\delta m^2_{W_\mu} &=& \frac{g_2^2}{64 \pi^4} m_{KK}^2\left[ - 39 T_6 (\xi) + (1+\xi^2) 35 \zeta(3) + \xi^2 (2 V_g (n/\xi) + V_s (n/\xi)) + \frac{47}{3} n^2 \pi^2 L\right]\,, \nonumber
\end{eqnarray}
for the vectors $n$-even.
In the $n$-even case, loops will also induce a mixing between the two tiers $(n,0)$ and $(0,n)$ proportional only to the divergent rotation contribution.
Such terms can be computed using the localised counter-terms, which are kinetic terms for the vector bosons~\cite{Cacciapaglia:2011hx}.
The off-diagonal contribution to the mass can be written as:
\begin{equation}
\delta m^2_{(n,0)-(0,n)} = \frac{g^2 C(r)}{64 \pi^4} m_{KK}^2\, \frac{1+\xi^2}{2} \; A_{(n,0)}\,,
\end{equation}
where the $A_{(n,0)}$ coefficients are the same as in the table (for the rotation only).
For the UED model, the off-diagonal terms are
\begin{eqnarray}
\left. \delta m^2_{G_\mu}  \right|_{(n,0)-(0,n)} &=& \frac{g_3^2}{64 \pi^4} m_{KK}^2\, \frac{1+\xi^2}{2} \left[24 n^2 \pi^2 L\right]\,, \nonumber \\
\left. \delta m^2_{B_\mu} \right|_{(n,0)-(0,n)} &=& \frac{g_1^2}{64 \pi^4} m_{KK}^2\, \frac{1+\xi^2}{2} \left[ - \frac{1}{3} n^2 \pi^2 L\right]\,, \\
\left. \delta m^2_{W_\mu}  \right|_{(n,0)-(0,n)} &=& \frac{g_2^2}{64 \pi^4} m_{KK}^2\, \frac{1+\xi^2}{2}\left[ \frac{47}{3} n^2 \pi^2 L\right]\,. \nonumber
\end{eqnarray}

\subsubsection*{$\xi$ dependence}

\begin{figure}[tb!]
\begin{center}
\includegraphics[width=16cm]{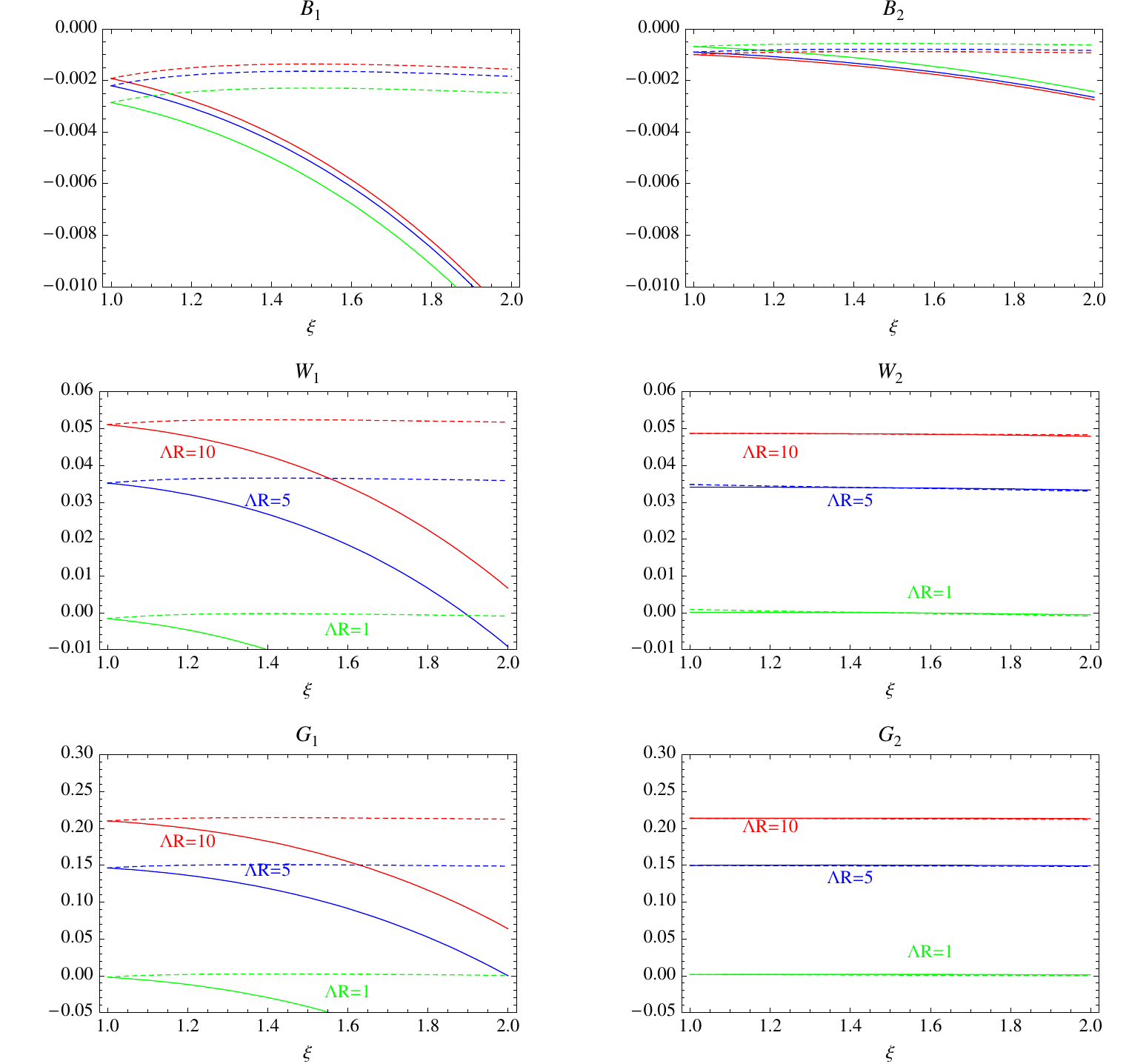}
\end{center}
\caption{\footnotesize $\delta m_g^2/m^2_g$ as a function of $\xi$: the three colours correspond to \textcolor{red}{$\Lambda R = 10$} ($L =4.6$), \textcolor{blue}{$\Lambda R = 5$} ($L=3.2$) and \textcolor{green}{$\Lambda R = 1$} ($L=0$); the solid line to tier $(n,0)$ while the dotted one to tier $(0,n)$; in the first column $n=1$, in the second $n=2$.}  \label{fig:dG}
\end{figure}

In Figure~\ref{fig:dG}, we plot the relative value of the loop corrections to the 3 gauge bosons of U(1) ($B$), SU(2) ($W$) and SU(3) ($G$) for the tiers $(1,0)$, $(0,1)$, $(2,0)$ and $(0,2)$: the plots only show the diagonal loop corrections, so the effect of cross-level mixing and of the Higgs VEV are not included.
The three colours correspond to different values of the cut-off $\Lambda$, in particular in green we show the correction for $\Lambda R = 1$, i.e. for vanishing divergent log ($L = 0$).
For the odd levels (on the left), the plots show a strong dependence on $\xi$ for the lightest mode $(1,0)$, while the dependence is much milder for the $(0,1)$ level.
The reason behind this behaviour is to be traced back to the strong $\xi$ dependence of $T_6$ (see the Appendix~\ref{app:functions} for more details) that contributes to the finite part of the correction, while the divergent part is independent on $\xi$. 
In fact, for $\xi>1$ the correction tends to more negative, due to the negative contribution of the $T_6$ term. 
In the $(0,1)$ case, dividing by the KK mass $4 \xi^2 m_{KK}^2$ cancels out the growth with $\xi$ of the finite part, thus the result is almost $\xi$-independent.
The finite part always plays a crucial role for the hypercharge, because of the smallness of the divergent contribution which is coming only from the Higgs loop.
For the even levels, this feature is even more obvious: the loop corrections are dominated by the divergent parts, thus the $\xi$-dependence is suppressed, except for the U(1) gauge vector $B$.
Nevertheless, the $\xi$ dependence is always milder than for the odd state.
This feature repeats itself for modes with larger $n$, both even and odd: the point is that the divergent term is proportional to $m_X^2$, while the finite contribution are only proportional to $1/R^2$ and therefore are suppressed by a factor $1/n^2$ with respect to the divergent ones.

\subsubsection*{Tier mixing}

For the odd modes, no mixing takes place, unless non negligible and asymmetric localised terms are present, therefore masses are well defined by equations \ref{gauge_corr1}. 
The localised terms always correspond to higher order operators (dimension 8 compared to the standard dimension 6 bulk terms)~\cite{Cacciapaglia:2011hx}, therefore we may expect them to be small compared to loop induced corrections.

In the (2,0)-(0,2) system, the mixings between the levels cannot be neglected because the off-diagonal terms are generated at loop level.
For simplicity, we will define the loop corrections to a generic $A_{(n,0)}$ and $A_{(0,n)}$ gauge vector (with $n$ even) as
$$
\left. \delta m^2_A \right|_{(n,0)} = n^2 m_{KK}^2 \delta_{(n,0)}\,, \quad \left. \delta m^2_A \right|_{(0,n)} = n^2 \xi^2 m_{KK}^2 \delta_{(0,n)}\,, \quad \left. \delta m^2_A \right|_{(n,0)-(0,n)} = n^2 \frac{1+\xi^2}{2} m_{KK}^2 \delta'_n\,.
$$
The two new mass eigenstates are defined as
\begin{eqnarray}
A_\mu^{n-} &=& \cos\varphi_n\; A_\mu^{(n,0)} - \sin\varphi_n\; A_\mu^{(0,n)}\,, \\
A_\mu^{n+} &=& \sin\varphi_n\; A_\mu^{(n,0)} + \cos\varphi_n\; A_\mu^{(0,n)}\,;
\end{eqnarray}
with masses
\beq
m_{2\pm}^2 = n^2 m_{KK}^2 \frac{1}{2} \left( 1 + \xi^2 + \delta_{(n,0)} + \xi^2 \delta_{(0,n)} \pm \Delta_n \right)
\eeq
where 
\beq
\Delta_n = \sqrt{(1+\xi^2)^2 (\delta_n')^2 + (\xi^2-1+ \xi^2 \delta_{(0,n)}- \delta_{(n,0)})^2}
\eeq
and the mixing angles $\varphi_n$ are a non-trivial functions of $\xi$:
\beq
\tan\varphi_n =\frac{\Delta_n -(\xi^2-1+ \xi^2 \delta_{(0,n)} - \delta_{(n,0)})}{ (1+\xi^2) \delta'_n}\,.
\eeq

Note that for $\xi^2-1 \gg \delta$'s we have
\beq
m^2_{n\pm} = n^2 m_{KK}^2 \times \left\{ \begin{array}{l} \xi^2 (1+\delta_{(0,n)}) \\ (1+\delta_{(n,0)}) \\ \end{array} \right.
\label{eq:mass_negligible_mix}
\eeq
with eigenstates $A_\mu^{n-} \sim A_\mu^{(n,0)}$ and $A_\mu^{n+} \sim A_\mu^{(0,n)}$.
 therefore the mixing will be negligible and we can and the two mass eigenvalues will be proportional to $m_{KK}$ and $\xi m_{KK}$.


In the degenerate case $\xi=1$, the loop corrections to the two modes are the same, $\delta_{(n,0)} = \delta_{(0,n)} = \delta_n$: this is true because the UED model does not contain bosons with $p_r = -1$ (let us stress again here that this is not a generic feature of the RP$^2$).
In this case, $\Delta_n = 2 \delta'_n$, and $\varphi_n = \pi/2$: the mixing is maximal and the mass eigenstates are given by
\begin{equation}
A_\mu^{n\pm} = \frac{1}{\sqrt{2}} \left( A_\mu^{(n,0)} \pm A_\mu^{(0,n)}\right)
\end{equation}
with masses
\begin{equation}
m^2_{n\pm} = n^2 m_{KK}^2 (1 + \delta_n \pm \delta'_n)\,.
\end{equation}
As the divergent part in $\delta_n$ and $\delta'_n$ have the same coefficient, it turns out that the fields $A_\mu^{n-}$ are insensitive to divergences: this can also be explained by the fact that their wave function vanishes on the singular points.


\subsection{Fermions}

In the fermion case, we can express the one-loop corrections from gauge loops to the masses as:
\begin{equation}
m_{(n,0)} \delta m_{(n,0)} = \frac{g^2 C_2(r_f)}{64 \pi^4} m_{KK}^2 \; F_{(n,0)}\,,
\end{equation}
where the fermion is in the representation $r_f$ and the group theory factor $C_2 (r_f)$ is replaced by the charge squared for U(1) loops.
For scalar loops via Yukawa interactions, it is enough to replace $g^2 C_2 (r_f) \to y_f^2$, where $y_f$ is the effective 4D Yukawa coupling.
The coefficients $F_{(n,0)}$ are summarised in the following table:
\begin{center}
\begin{tabular}{ll|c|c|c|c|}
\multicolumn{2}{c|}{$F_{(n,0)}$} & torus & glide ($p_g \times$) & glide' ($p_g p_r \times$) & rotation ($p_r\times$) \\
 \hline
 \multirow{2}{*}{gauge loops} & $n$-even & \multirow{2}{*}{$0$} & $0$ & \multirow{2}{*}{$\xi^2 \left( 7 \zeta(3) + F_g (n/\xi) \right)$} & \multirow{2}{*}{$4 n^2 \pi^2 L$}\\
 &  $n$-odd & & $14 \zeta(3)$ &  & \\
 \hline
  \multirow{2}{*}{scalar loops} & $n$-even & \multirow{2}{*}{$0$} & $0$ & \multirow{2}{*}{$\xi^2 \left( \frac{7}{2} \zeta(3) + F_s (n/\xi) \right)$} & \multirow{2}{*}{$\frac{1}{2} n^2 \pi^2 L$}\\
 & $n$-odd & & $7 \zeta(3)$ & & \\
 \hline
 \end{tabular} \end{center} 
where the small and finite functions $F_{g,s}$ are shown in Appendix~\ref{app:functions}, and the parities refer to the parity of the bosons in the loop (gauge or Higgs). The only difference between even and odd $n$ is in the contribution of the glide, which is absent for even tiers.

In the UED model all parities are $(+,+)$, thus the mass corrections for a generic fermion is
\begin{eqnarray}
\left.\delta m_{(n,0)} \right|_{n-even} &=& \frac{m_{KK}}{64 \pi^4 n} \left[ \left( \sum_{gauge} g_i^2 C_2(r_f) \right) \left( \xi^2 7 \zeta(3) + \xi^2 F_g (n/\xi) + 4 n^2 \pi^2 L\right) \right. \nonumber \\ 
& & \left. + y_f^2 \left(  \frac{\xi^2}{2} 7 \zeta(3) + \xi^2 F_s (n/\xi) +\frac{1}{2} n^2 \pi^2 L\right) \right]\,, \\
\left. \delta m_{(n,0)} \right|_{n-odd} &=& \frac{m_{KK}}{64 \pi^4 n} \left[ \left( \sum_{gauge} g_i^2 C_2(r_f) \right) \left( (2+\xi^2) 7 \zeta(3) + \xi^2 F_g (n/\xi) + 4 n^2 \pi^2 L\right) \right. \nonumber \\ 
& & \left. + y_f^2 \left( \frac{2+\xi^2}{2} 7 \zeta(3) + \xi^2 F_s (n/\xi) +\frac{1}{2} n^2 \pi^2 L\right) \right]\,.
\end{eqnarray}

The $n$-even fermions mix via localised counter-terms: in the fermionic case, there are both kinetic terms and bulk mass terms (i.e. terms with a derivative along the extra co-ordinates), however the off-diagonal mass corrections can be cast in a simple way
\begin{eqnarray}
\delta m_{(n,0)-(0,n)} &=& \frac{g^2 C_2(r_f)}{64 \pi^4 n} m_{KK} \; F_{(n,0)}\,, \\
\delta m_{(0,n)-(n,0)} &=& \xi\, \delta m_{(n,0)-(0,n)}\,,
\end{eqnarray}
where only the divergent terms in $F_{(n,0)}$ contribute.
Note that the term proportional to $\xi$ may be $\delta m_{(n,0)-(0,n)}$, depending on the parity of the field: in fact, the off diagonal mass terms are proportional to the mass of the fermion component which is odd under the rotation $r$, the reason being that the non vanishing localised term must be proportional to the derivative of its wave function.
Therefore, with the standard mass definition in 4 D $m\, \bar{\psi} \psi$, where $\psi$ is a generic Dirac spinor describing the fermion KK modes, then $\delta m_{(0,n)-(n,0)}$ is proportional to $\xi$ for a fermion with left-handed zero mode, while the other situation occurs for fermions with right-handed zero modes.

For a UED fermion:
\begin{equation}
\delta m_{(n,0)-(0,n)} = \frac{m_{KK}}{64 \pi^4 n} \left[ 4 \sum_{gauge} g_i^2 C_2(r_f)  + \frac{1}{2} y_f^2 \right]\, n^2 \pi^2 L \,.
\end{equation}

\subsubsection*{$\xi$ dependence}

\begin{figure}[tb!]
\begin{center}
\includegraphics[width=16cm]{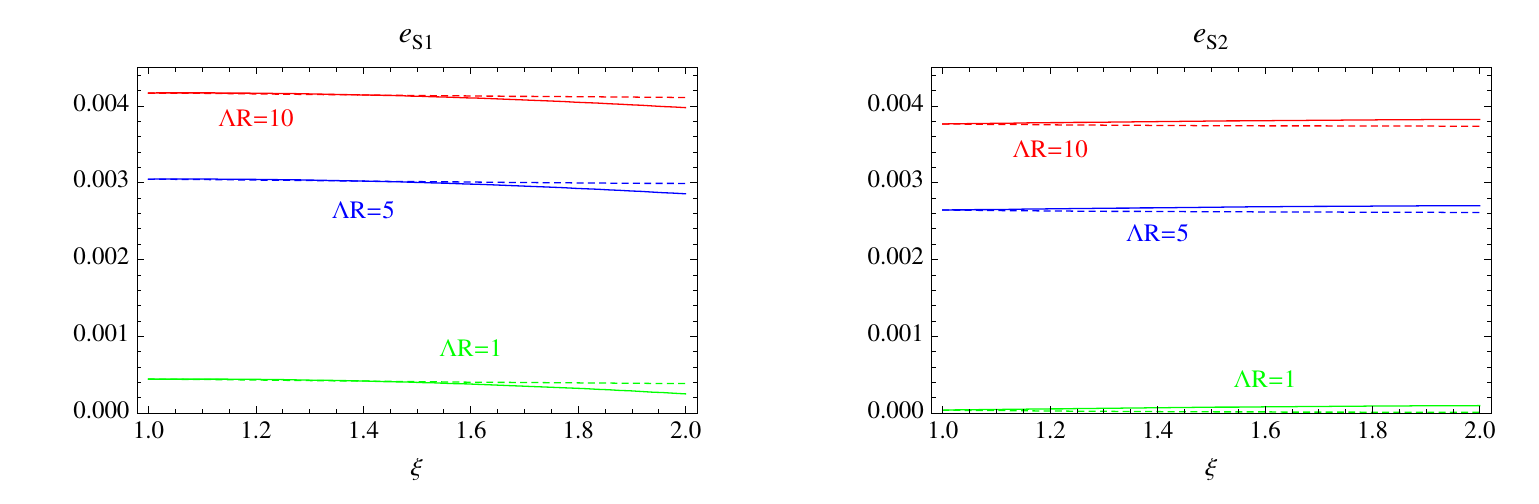}
\end{center}
\caption{\footnotesize $\delta m_f/m_f$ as a function of $\xi$ for singlet leptons: the three colours correspond to \textcolor{red}{$\Lambda R = 10$} ($L =4.6$), \textcolor{blue}{$\Lambda R = 5$} ($L=3.2$) and \textcolor{green}{$\Lambda R = 1$} ($L=0$); the solid line to tier $(n,0)$ while the dotted one to tier $(0,n)$; in the first column $n=1$, in the second $n=2$. The plots show a mild dependence on $\xi$ in both cases.}  \label{fig:dF}
\end{figure}

In Figure~\ref{fig:dF}, we show the relative mass correction $\delta m/m$ for the fermions.
In the plot we show the results for the singlet leptons $E$, as the behaviour is very similar for other fermions.
The plot shows a very mild dependence on $\xi$ for both even and odd tiers: this behaviour is to be understood in relation to the absence of bulk contributions to the mass proportional to $T_6 (\xi)$.
The masses, therefore, are always dominated by the divergent terms, while the finite contributions are always very small (green lines in the plot).
The plot also shows that the relative mass correction is very similar for the two modes $(n,0)$ and $(0,n)$.

\subsubsection*{Tier mixing}

Similarly to the bosonic case, the only mixing generated at loop level occurs for the even modes.
For a generic value of $\xi$, one can diagonalise the mass matrix by use of two rotation matrices, acting on the two chiralities of the fermion (labelled by $L$ and $R$ for left- and right-handed chirality respectively): 
\beq
\psi_{n-,L/R} &=& \cos \varphi_{n,L/R}\; \psi_{(n,0),L/R} - \sin \varphi_{n,L/R}\; \psi_{(0,n),L/R}\,,\\
\psi_{n+,L/R} &=& \sin \varphi_{n,L/R}\; \psi_{(n,0),L/R} +\cos \varphi_{n,L/R}\; \psi_{(0,n),L/R} \,.
\eeq

For a KK tower with a left-handed zero mode, the mass corrections can be written as
$$
\begin{array}{c}
\left. \delta m_\psi \right|_{(n,0)} = n m_{KK} (1+\delta_n)\,, \quad \left. \delta m_\psi \right|_{(0,n)} = n \xi m_{KK} (1+\delta_n)\,, \\
\left. \delta m_\psi \right|_{(n,0)-(0,n)} = n m_{KK} \delta'_n\,, \quad \left. \delta m_\psi \right|_{(n,0)-(0,n)} = n \xi m_{KK} \delta'_n\,; \end{array}
$$
where we have assumed that the loop corrections for the two tiers are the same, $\delta_{(n,0)} \sim \delta_{(0,n)} = \delta_n$.
In this approximation, the mixing angles $\varphi_{n,L/R}$ and the mass eigenstates $m_{2\pm}^2$ and are given by 
\beq
\tan\varphi_{n,L} &=& \frac{\Delta_n^F - (\xi^2-1) \left((1+\delta_n)^2 - {\delta_n'}^2 \right)}{2 (1+\xi^2) \delta'_n (1+\delta_n)}\,, \\
\tan\varphi_{n,R} &=& \frac{\Delta_n^F - (\xi^2-1) \left((1+\delta_n)^2 + {\delta_n'}^2 \right)}{4 \xi \delta'_n (1+\delta_n)}\,;
\eeq
and
\beq
m_{2\pm}^2 = \frac{n^2}{R^2} \frac{1}{2} \left( (1+\xi^2) \left((1+\delta_n)^2 + {\delta'_n}^2\right) \pm \Delta_n^F \right)\,,
\eeq
where
\beq
\Delta_n^F = \sqrt{(1+\xi^2)^2 \left( (1+\delta_n)^2 + {\delta'_n}^2\right)^2 - 4 \xi^2 \left((1+\delta_n)^2 - {\delta'_n}^2\right)^2}\,.
\eeq
As before, for $\xi-1 \ll \delta'_n$, the two tiers decouple.
For a KK tower with right-handed zero mode, the only difference is that
$$
\begin{array}{c}
\left. \delta m_\psi \right|_{(n,0)-(0,n)} = n \xi m_{KK} \delta'_n\,, \quad \left. \delta m_\psi \right|_{(n,0)-(0,n)} = n m_{KK} \delta'_n\,; \end{array}
$$
and the two mixing angles are exchanged, while the mass eigenvalues are unaffected.

In the degenerate radii case, $\xi=1$, the situation is similar to the bosonic case: the mass corrections for the two degenerate tiers are exactly the same, and the mixing between mass eigenstates becomes maximal. 
In fact, for $\xi=1$, $\Delta_n^F = 4 (1+\delta_n) \delta'_n$, and $\varphi_{n,L} = \varphi_{n,R} = \pi/2$.
The mass eigenstates are the sum and difference of the two modes, with masses
\begin{equation}
m_{n\pm} =n m_{KK} (1+\delta_n \pm \delta'_n)\,.
\end{equation}
As in the bosonic case, the mode $\psi_{n-}$ only received finite corrections due to the fact that the divergent part in $\delta_n$ and $\delta_n'$ are the same.

\subsection{Scalars (Higgs field)}

In UED models the Brout-Englert-Higgs mechanism is implemented in the same way as in the Standard Model, i.e. by introducing a bulk scalar doublet which, due to a properly engineered potential, acquires a vacuum expectation value (VEV) and gives mass to $W$, $Z$ and fermions.
Here we will focus on the simplest possibility that the Higgs has a zero mode, thus parities $(+,+)$, and that the VEV is generated  by a negative bulk mass.
Denoting by $\Phi_H$ the bulk scalar, the tree level potential is
\begin{equation}
\mathcal{L}_{\rm bulk} = \left( D_\mu \Phi_H \right)^\dagger D^\mu \Phi_H - m_B^2 \Phi_H^\dagger \Phi_H + \frac{\lambda_B}{2} \left( \Phi_H^\dagger \Phi_H \right)^2\,,
\end{equation}
where $m_B$ is the bulk mass, $\lambda_B$ is the quartic coupling; note that the quartic operator has dimension in mass 8 (as the scalar field has dimension 2), therefore the coupling $\lambda_B$ has dimension $-2$.
This fact may rise the problem that, in order to have a heavy enough Higgs boson, we need a higher order operator with unnaturally large coupling: here we will brush this issue aside, as the Higgs sector of this model should just be taken as a toy model because it has more serious issues than the SM Higgs itself.

Loop corrections will also affect such potential and their structure is similar to the gauge bosons: there are finite contributions from the torus and glides which are almost mode-independent, the only difference coming from the small $n$-dependent contributions generated by the glide', while the log divergent terms are proportional to the mass of the mode.
The loops also contain quadratically divergent contributions which correspond to a localised mass.
The mode-independent corrections can be reabsorbed in the bulk mass, which is a free parameter, therefore we will neglect them in the following: the only contributions that affect the spectrum are thus the localised contributions that can be parametrised in terms of the counter-term Lagrangian~\cite{Cacciapaglia:2011hx}
\begin{multline}
\Lambda^2 \mathcal{L}_{\rm loc} = c_1 \left( D_\mu \Phi_H \right)^\dagger D^\mu \Phi_H + c_4' \left( \partial_4^2 \Phi_H^\dagger\, \Phi_H + \Phi_H^\dagger\, \partial_4^2 \Phi_H \right)  \\ + c_5' \left( \partial_5^2 \Phi_H^\dagger\, \Phi_H + \Phi_H^\dagger\, \partial_5^2 \Phi_H \right) - \delta m^2 \Phi_H^\dagger \Phi_H\,,
\end{multline}
with coefficients
\begin{eqnarray}
\frac{c_1}{\pi^2 \Lambda^2 R_4 R_5} &=& - 3 \left( g_1^2 + 3 g_2^2 \right) \frac{n^2 L}{64 \pi^2}\,,\\
\frac{c'_{4,5}}{\pi^2 \Lambda^2 R_4 R_5} &=& - \frac{5}{8} \left( g_1^2 + 3 g_2^2 \right) \frac{n^2 L}{64 \pi^2}\,.
\end{eqnarray}
Note that $\delta m^2$, which is a brane localised mass, is a free parameter, and that the radiative corrections do not include loops proportional to the quartic coupling.
Furthermore, the localised mass term is a dimension 6 operator (4 from the two scalar fields and 2 from the localising delta functions), therefore it must be considered as a new tree level parameter: not that the coefficient $\delta m^2/\Lambda^2$ is dimension-less.

In order to understand the spectrum, we need first to study the effective potential for the zero mode, which corresponds to the SM Higgs boson.
In terms of the bulk and localised Lagrangians, the potential for the zero mode is
\begin{equation}
V(\phi_{(0,0)}) = (m_B^2 + m_b^2 ) \phi_{(0,0)}^\dagger \phi_{(0,0)} + \frac{1}{2} \lambda (\phi_{(0,0)}^\dagger \phi_{(0,0)} )^2\,,
\end{equation}
where $\lambda = \frac{\lambda_B}{4 \pi^2 R_5 R_6}$ is the effective 4D quartic coupling and $m_b^2 = \frac{\delta m^2}{2 \pi^2 \Lambda^2 R_5 R_6}$ is the contribution from the localised mass.
After the usual expansion
\begin{equation}
\phi_{(0,0)} = \frac{1}{\sqrt{2}} \left( \begin{array}{c}
0 \\ v + H \end{array} \right)\,,
\end{equation}
we obtain that the SM VEV $v$ and the Higgs mass are given by
\begin{equation}
m_H^2 = - 2 (m_B^2 + m_b^2) = \lambda v^2\,. \label{eq:Higgsmass}
\end{equation}
In the following, we will consider $m_b$ a free parameter and replace $m_B$ with $m_H$ (a measured quantity) in all expressions by use of the relation in Eq.~\ref{eq:Higgsmass}.
Note also that we can expect $m_b$ to be below the TeV scale in order not to introduce a severe fine tuning in the Higgs mass: this is the only concession to naturalness we will have in this model.
Note also that both bulk mass $m_B$ and localised mass $m_b$ suffer from quadratically divergent loop corrections: therefore, in UED models, the naturalness problem is doublet compared to the 4D SM.

For the $(n,0)$ mode, the Lagrangian contains the following mass term, including the contribution of the VEV:
\begin{equation}
- \left( n^2 m_{KK}^2 (1-\delta_\phi) + m_B^2 + 2 m_b^2 + \frac{\lambda v^2}{2} \right) \phi^\dagger_{(n,0)} \phi_{(n,0)} - \lambda v^2 \left( \mbox{Re} \phi^0_{(n,0)} \right)^2\,,
\end{equation}
where $\phi^0$ is the down component of the doublet and
\begin{equation}
\delta_\phi =  - \frac{c_1 + 2 c'_5}{\pi^2 \Lambda^2 R_4 R_5} = \frac{17}{4} (g_1^2 + 3 g_2^2) \frac{L}{64 \pi^2}
\end{equation}
contains the log-divergent contributions.
We can now expand the Higgs doublet as
\begin{equation}
\phi_{(n,0)} = \left( \begin{array}{c}
\pi^+_{(n,0)} \\
\frac{H_{(n,0)} + i \pi^0_{(n,0)}}{\sqrt{2}}
\end{array} \right) 
\end{equation}
whose components have mass
\begin{equation}
m^2_{\pi_{(n,0)}} = n^2 m_{KK}^2 (1-\delta_\phi) + m_b^2\,, \qquad m^2_{H_{(n,0)}} = m^2_{\pi_{(n,0)}} + m_H^2\,.
\end{equation}
The fields $\pi^{\pm,0}$ are the would-be goldstone bosons that would provide the longitudinal polarisation to the massive gauge bosons if the Higgs VEV were the only source of mass.
However, for the $(n,0)$ modes, the main contributions to the mass comes from the extra dimension, and the role of the goldstone bosons is played by the extra polarisations of the vectors.
The Higgs VEV will in general mix the $\pi^{\pm,0}$ with the gauge scalars: one effect of such mixing is to introduce a new correction to the mass of the physical states $s^{\pm,0}_{(n,0)}$.
The correct masses are therefore given by:
\begin{equation}
m^2_{s^\pm_{(n,0)}} = m^2_{\pi_{(n,0)}}  + m_W^2\,, \qquad m^2_{s^0_{(n,0)}} = m^2_{\pi_{(n,0)}}  + m_Z^2\,.
\end{equation}
For the $(0,n)$ modes, the same formulas apply, with the simple substitution $m_{KK} \to \xi m_{KK}$.

The localised counter-terms will also generate a mixing between states with different KK number.
An important consequence is that the VEV spreads to other KK states and not only to the zero mode.
Such corrections are can be relevant, as they will generate mixing terms between zero modes and heavy states, thus generating corrections to electroweak observables at tree level.
Among others, mixing between the $(n,0)$ and $(0,n)$ states will be generated.
As the quartic coupling is in the bulk, the off-diagonal terms will be the same for the 4 Higgs states:
\begin{equation}
\delta m^2_{\pi, (n,0)-(0,n)} = - \frac{1+\xi^2}{2} n^2 m_{KK}^2 \delta_\phi + 2 m_b\,.
\end{equation}

\subsubsection*{$\xi$ dependence}

The masses of the KK Higgs bosons are only sensitive to the localised divergent contribution of the loops, which are proportional to the masses of the modes.
A very mild $\xi$ dependence remains via the small finite and $n$-dependent terms appearing with the glides, however such terms are negligibly small.
The only term that will introduce a $\xi$ dependence in the relative mass contribution is the localised mass $m_b$, which is a constant: this contribution can be dealt with on the same footing as the Higgs VEV one, and not like a loop contribution.

\subsubsection*{Tier mixing}

The structure of the tier mixing for the Higgs bosons is similar to the gauge boson ones, therefore he same general formulas can be used in this case.

In the degenerate radii case $\xi=1$, the diagonal masses are the same, therefore as before the mass eigenvalues are given by sum and difference with eigenvalues
\begin{eqnarray}
m_{\pi,n+}^2  &=& m_{\pi,(n,0)}^2 +  \delta m^2_{\pi, (n,0)-(0,n)} = n^2 m_{KK}^2 (1- 2 \delta_\phi) + 3 m_b^2\,,  \\
m_{\pi,n-}^2  &=& m_{\pi,(n,0)}^2 -  \delta m^2_{\pi, (n,0)-(0,n)} = n^2 m_{KK}^2 - m_b^2\,.
\end{eqnarray}
As before, $m_{\pi, n-}$ is insensitive to the logarithmic divergences, however it depends on the localised mass $m_b$.
The contribution of the Higgs VEV to the masses of $H$, $s^0$ and $s^\pm$, are given by the same formulas as in the non-degenerate case.

\section{Numerical spectra and Monte Carlo implementation}
\label{sec:num}

We now turn our attention specifically to a UED incarnation of the Real Projective plane, where a 6D field is associated to each SM field and the SM Lagrangian is extended to 6D.
Both tiers $(2,0)$ and $(0,2)$ contain a massive vector boson for each SM gauge boson, a Dirac fermion for each chiral SM fermion and 4 scalars ($H$, $S^0$ and $S^\pm$) corresponding to the Higgs field.

Neglecting higher order operators, the model has 4 free parameters: the mass scale $m_{KK} = 1/R_4$, the asymmetry factor $\xi = R_4/R_5$, the cut-off $\Lambda$ and the localised Higgs mass $m_b^2$.
The cut-off enters logarithmically the loop corrections to masses and the loop induced KK momentum violating couplings.
Naive dimensional analysis suggests that $\Lambda \sim 5 \div 10 \cdot m_{KK}$, and in the following we will always fix $\Lambda = 10\, m_{KK}$, a situation in which the mass splittings are maximal.
The localised Higgs mass contribution $m_b^2$ only enters the mass of the Higgs resonances which are not very relevant for the LHC phenomenology being rarely produced and difficult to detect.
For simplicity, we will fix $m_b = 0$ GeV from now on (note that $m_b^2$ may be negative and drive the electroweak symmetry breaking in the SM Higgs sector).

\begin{table}[t!]
\begin{center}
\begin{tabular}{c|c|c|c|c||c|c|}
$m_{KK}$ & \multicolumn{4}{c||}{500} & \multicolumn{2}{c|}{calcHEP 500}\\
$\xi$ & 1 & 1.1 &1.2 & 1.5 & sym & asym \\
\hline
\multirow{2}{*}{$e_S^{(2)}$} & 1000.0 & {\bf 1003.6} & {\bf 1003.7} & {\bf 1003.8} &  & 1003.8 \\
                                             & {\bf 1007.5}  & 1104.3 &1204.6 & 1505.6 & 1007.5 & \\
\hline
\multirow{2}{*}{$e_D^{(2)}$ and $\nu^{(2)}$} &  1000.1 & {\bf 1009.2} & {\bf 1009.8} & {\bf 1010.1} &  & 1010.3\\
                                             & {\bf 1020.6} & 1112.5 & 1213.0 & 1515.7 &1020.6&\\
\hline
\multirow{2}{*}{$d_S^{(2)}$ and $b_S^{(2)}$}  & 1000.5 & {\bf 1028.2} & {\bf 1035.9}  & {\bf 1041.9} &  & 1047.5\\
                                              & {\bf 1095.6} & 1172.8 & 1269.8 & 1578.4 &1094.4&\\
\hline
\multirow{2}{*}{$u_S^{(2)}$} & 1000.5 & {\bf 1028.5} & {\bf 1036.6}  & {\bf 1042.8} &  & 1048.7\\
                                             & {\bf 1098.1} & 1175.0 & 1272.0 & 1580.6 &1096.9&\\
\hline
\multirow{2}{*}{$u_D^{(2)}$ and $d_D^{(2)}$} & 1000.6 & {\bf 1030.6} & {\bf 1040.4}  & {\bf 1048.5}  &  & 1056.6\\
                                             & {\bf 1113.7} & 1189.4 & 1285.3 & 1594.6 &1112.5&\\
\hline
\multirow{2}{*}{$b_D^{(2)}$} & 1000.8 & {\bf 1031.6} & {\bf 1042.3} & {\bf 1051.4} &  & 1060.4\\
                                             & {\bf 1121.3}  & 1196.6 & 1292.0 & 1601.5 &1120.1&\\
\hline
\multirow{4}{*}{$t_{S/D}^{(2)}$} & 1015.9 & {\bf 1044.5} & {\bf 1053.3} & {\bf 1060.4}  & & 1067.0\\
                                                  & 1016.0 & {\bf 1046.4} & {\bf 1056.9} & {\bf 1065.9}&&1074.8\\
                                                  & {\bf 1119.4} & 1194.9 & 1290.4 & 1597.0 & 1118.2 & \\
                                                  & {\bf 1135.0} & 1209.4 & 1303.9 & 1611.1 &1133.8&\\
\hline
\hline
\multirow{4}{*}{$A^{(2)}/Z^{(2)}$} & {\bf 999.9} & {\bf 1000.3} & {\bf 1000.2}  & {\bf 1000.0} & 1000.2 & 1000.3\\
                                                          & 1000.2  & {\bf 1021.7} & {\bf 1024.0} & {\bf 1025.2}&&1027.3\\
                                                          & 1004.8  & 1100.4 & 1200.5 & 1500.6&  & \\
                                                         & {\bf 1049.8} & 1135.0 & 1235.5 & 1542.1 &1050.6&\\
\hline
\multirow{2}{*}{$W^{(2)}$}  &  1004.5 & {\bf 1021.5} & {\bf 1023.8} & {\bf 1025.0}&  & 1027.2 \\
                                           & {\bf 1049.8} & 1134.4 & 1234.2 & 1539.4 &1050.5&\\
\hline
\multirow{2}{*}{$G^{(2)}$} & 1003.0  & {\bf 1038.8} & {\bf 1055.3} & {\bf 1069.9} &  & 1100.4\\
                                          & {\bf 1192.1} & 1266.0 & 1359.2 & 1672.8 &1191.8&\\
\hline
\hline
\multirow{2}{*}{$H^{(2)}$} & {\bf 963.7}  & {\bf 980.5} & {\bf 982.7} & {\bf 983.9} & 963.7 & 986.0\\
                                          &  1008.8 & 1088.1 & 1183.1 & 1473.8 &  &\\
\hline
\multirow{2}{*}{$S_0^{(2)}$} &  {\bf 959.9}  & {\bf 976.7} & {\bf 978.9} & {\bf 980.2} & 959.9 & 982.3\\
                                          & 1004.1 & 1084.7 & 1180.0 & 1471.0 & &\\
\hline
\multirow{2}{*}{$S^{(2)\pm}$} & {\bf 958.9}  & {\bf 975.7} & {\bf 978.0} & {\bf 979.2} & 958.9 & 981.3\\
                                          & 1003.2 & 1083.8 & 1179.2 & 1470.7 & &\\
\hline
 \end{tabular} 
\caption{\footnotesize Spectra of the even tiers $(2,0)$ and $(0,2)$ for $m_{KK} = 500$ GeV: in the first 4 columns the exact results including one loop and Higgs VEV contributions; in the latter 2 columns the approximate results implemented in CalcHEP for the degenerate (symmetric) case and asymmetric one. In bold, we highlighted the $2_+ = (2,0) + (0,2)$ combination in the case $\xi=1$, and the states whose mass is proportional to $m_{KK}$ in the other cases: the bold numbers are the ones to be compared with the CalcHEP values. For the Higgses, we set $m_{b} = 0$ and $m_H = 125$ GeV.} \label{tab:spectrum}
\end{center} \end{table}

The physical spectrum of the even tiers, $(2,0)$ and $(0,2)$, depends in a non trivial way on the ratio of the radii $\xi$.
Furthermore, when the difference between the two radii is of the same order as the loop induced corrections to the masses, significant mixing between the two tiers occurs. 
To better illustrate this behaviour we focused on a benchmark point with $m_{KK} = 500$ GeV: in Table~\ref{tab:spectrum} we show, in the first 4 columns, the mass eigenstates of the even tiers for 4 values of $\xi = 1$ (symmetric case), $\xi = 1.1$, $1.2$ and $1.5$. 
For each particle, there are two values which correspond to the mass eigenstates of the mixed system $(2,0)$-$(0,2)$.
For the neutral gauge bosons, there is also a mixing induced by the Higgs VEV between the hypercharge $B$ and the neutral SU(2) $W^3$ gauge bosons, thus the 4 mass eigenstates of the complete system are quoted.
Similarly, for the top, the Yukawa coupling induces a mixing between the singlet and doublet top resonances (we here neglect the effect of the other Yukawa couplings), thus 4 eigenvalues are quoted.
To understand the spectrum, let's consider first the case of the singlet charged leptons $e_S^{(2)}$: for increasing $\xi$, we can clearly see the appearance of two states whose mass is proportional to $m_{KK}$ and $\xi m_{KK}$ respectively.
Moreover, for large $\xi$, the relative correction is the same for the two states, i.e. $1.0038 \cdot m_{KK}$ and $1.0038 \cdot \xi m_{KK}$.
This shows a clear separation between the two states $(2,0)$ and $(0,2)$, and also that the correction is approximately independent on $\xi$.
The largest loop corrections arises for the gluon and it amounts to about 10\% of the mass: this means that for $\xi \gtrsim 1.1$ the two states should already be separated.
In the symmetric case, the mixing is maximal and we observe the separation of a set of states with very small corrections and a set with large corrections (the latter being highlighted in bold in the table): the two sets correspond to the difference and sum mass eigenstates.
The former is only sensitive to finite corrections, therefore the mass splittings are very small, while the latter is affected by twice the log divergent contribution.
In fact, the relative correction is approximately doubled compared to the asymmetric case: for the $e_S^{(2)}$, for instance, the correction is $0.75\%$.
This separation is not a general property of the RP$^2$, but, as discussed in the previous sections, it derives from the fact that there are no bosonic fields with parity $p_r = -1$ in the model.
Another important consequence of this accidental occurrence is that the loop induced couplings between states in the tiers $(2,0)$ and $(0,2)$ to SM particles are the same.
Therefore, only the sum mass eigenstates (the one with the largest mass splitting) couples to SM particles at one loop level.
In other words, only the states with large splittings will be able to decay into a pair of SM states.
The states with small splittings can decay neither into SM states at one loop level, nor into a pair of odd states due to the small mass corrections.
This is important for the phenomenology, because this kind of states will only be allowed to chain decay into the lightest state of their kind, emitting very soft SM particles.
The lightest state, which is a neutral vector boson, is not protected by any symmetry thus it will eventually decay into SM states via higher order operators.
This may lead to interesting resonant final states, however the branching rations and lifetime will depend on unknown parameters.
For the time being, we can safely assume that the lifetime of the neutral vector boson is large enough for this particle to escape the detector without leaving any signatures.
Furthermore, such states will not play any significant role in the phenomenology of the odd states, as they cannot decay into them and they cannot participate as s-channel resonance in their pair production.

From this discussion, we can therefore identify three distinct physical regimes:
\begin{itemize}
\item[-] decoupling regime (asymmetric radii): when $\xi$ is above one, the two sets $(n,0)$ and $(0,n)$ decouple, with the latter being heavier. The mass corrections, relative to the tree level mass of the tier, are approximately the same.
\item[-] degenerate regime (symmetric radii): when $\xi = 1$, in the UED model, the $(2,0)$-$(0,2)$ system splits into two subsets, of which the one with larger loop corrections is phenomenologically interesting. The odd states are exactly degenerate and have identical couplings.
\item[-] intermediate case: in this case, the mixing is not trivial. However, it only happens when the radii are different but close enough, therefore in a very small region of the parameter space.
\end{itemize}

\subsection{Monte Carlo implementation}

We implemented this model in the FeynRules~\cite{feynrules} package of Mathematica, which allows to interface the model with Monte Carlo generators like CalcHEP, MadGraph4 and 5.
The implementation should contain the full tiers $(1,0)$, $(0,1)$, $(2,0)$, $(0,2)$ with $\xi$-dependent mass corrections and mixing between the two even tiers.
The tier $(1,1)$ may also give rise to interesting phenomenology, however there is a dependence on the unknown coefficients of the localised Lagrangians, which allow for the direct decays into a SM pair: an example of a spectacular 4-top signal has been studied in~\cite{4tops}.
In the following we will not consider this tier any further.
The full implementation would contain a huge number of new states and interactions.
Following the above discussion of the three physical regimes, however, we can simplify the model by considering two separate limits: the case of asymmetric radii and the degenerate case.

\paragraph{Asymmetric regime}
\

For asymmetric radii, $\xi \gtrsim 1.1$, the two sets $(n,0)$ and $(0,n)$ decouple because the mixing between the even states are negligible.
Furthermore, the relative mass corrections are nearly independent on $\xi$: the model reduces to two uncorrelated sets, characterised my $m_{KK} = 1/R$ and $m_{KK} = \xi/R$.
To simplify the implementation, therefore, we can consider a model with only one set of odd $(1,0)$ states and one set of even $(2,0)$ modes (for convenience, we can label them with a $(1)$ and $(2)$ dropping the zero).
The heavier states $(0,1)$ and $(0,2)$ can be described by the same simplified model with rescaled mass scale $m_{KK}$.

Thus, in order to study the phenomenology of a model with $m_{KK}$ and $\xi$, it suffices to generate events with mass scales $m_{KK}$ and $\xi\, m_{KK}$ and sum the two results.

\paragraph{Degenerate (symmetric) regime}
\

In the symmetric case, the model contains two degenerate and identical odd tiers, while the even tiers consist of one (the sum) that couples to SM states and has larger mass splittings, and one (the difference) that does not. 
The latter is not relevant for the phenomenology we are interested in.
To simplify the implementation, we can study a model that includes one of the two odd states, and the even states with larger mass splittings: the particle content is therefore the same as in the asymmetric simplified model!
In the mass corrections, we need to supplement a factor of 2 in front of the log divergent contributions to the even masses only.
Furthermore, when studying the phenomenology of the odd tier, we need to multiply the results by a factor of 2 to take into account the presence of two degenerate and identical tiers.
The couplings of the odd tier will also be affected by the change of basis: here we are interested uniquely in the phenomenology of even and odd tiers.
The processes we will consider are:
\begin{itemize}
\item[-] pair production or annihilation of odd tiers, which are sensitive to the even tier via the couplings $(2,0) (1,0) (1,0)$ and $(0,2) (0,1) (0,1)$ in diagrams with a resonant s-channel even state. In the mass eigenstate, the two couplings pick up a factor of $1/\sqrt{2}$. However, this coupling will also enter the decays of the even state into a pair of odd states: in our simplified model which includes only one odd state, this partial width would be underestimated by a factor of two. To compensate for this, and have the correct branching ratios, we do not include the $1/\sqrt{2}$ in the coupling. Therefore, when computing resonant production of the odd states, or annihilation via a resonant even state, it is intended that the sum over the two degenerate odd tiers is implicitly done.
\item[-] pair production of the even states are sensitive to the couplings $(0,0) (2,0) (2,0)$ and $(0,0) (0,2) (0,2)$, which are equal at three level. In the mass eigenstates, no extra factor is generated. These couplings will also generate decays $(2) \to (2)' + SM$.
\item[-] single production of the even state: this process is sensitive to the loop induced couplings $(2,0) (0,0) (0,0)$ and $(0,2) (0,0) (0,0)$, which have the same coefficient. In the mass eigenstate, the coupling has an extra factor $\sqrt{2}$. This coupling will also enter the decays $(2) \to SM$.
\end{itemize}

We have implemented the two simplified models in FeynRules, and interfaced them with both MadGraph and CalcHEP.
As the field content and couplings are the same, we have implemented a single model and introduced a parameter that distinguishes the two cases by adding the appropriate $\sqrt{2}$ factors as described above.
In the implementation of the mass corrections, we simply use the formulas for $\xi = 1$, which are formally valid only in the symmetric case, relying on the fact the mass corrections show a very mild dependence on $\xi$.
In the last two columns in Table~\ref{tab:spectrum}, we show the spectra as computed in the CalcHEP implementation of the model.
We see that the approximate masses closely reproduce both the asymmetric and symmetric cases.
The model also contains the full set of loop induced couplings, as computed in Ref.~\cite{Cacciapaglia:2011hx}, and the full one loop masses for both even and odd states.

\section{Estimates of LHC 2011 bounds}
\label{sec:LHC}

The LHC experiments have collected an impressive amount of data, 5 fb$^{-1}$ per experiment~\footnote{Here we focus on the general purpose ATLAS and CMS detectors.} in 2011 at a centre of mass energy of 7 TeV and a similar amount so far at 8 TeV in 2012 with a forecast integrated luminosity of 30 fb$^{-1}$ collected by the end of 2012, before the 2 year technical stop.
Even though no search specific to this model have been performed, many of the searches designed for supersymmetry or other exotic signals and models can be used to pose severe bounds on the RP$^2$.

The lightest odd tiers will give rise to signatures similar to supersymmetry due to the decay chains ending inevitably into a stable neutral particle (missing transverse momentum). 
However, the spectrum of the tiers is rather compressed due to the smallness of the splittings: the most effective searches should be the ones looking for supersymmetric signals in events with jets and missing transverse energy where, in this case, the jets are coming mainly from initial state radiation.
A detailed simulation is essential in order to estimate the efficiency of the experimental cuts on the signal events, thus a study specific to this model is in progress~\cite{jetmet}.

In the following we will focus on the phenomenology of the even tiers $(2,0)$ and $(0,2)$: the main interest here is that, due to loop induced vertices, such states can decay directly into a pair of SM particles, thus giving rise to resonant signatures.
Even though the mass of the even tiers is roughly double compared to the lightest odd tiers, thus the productions cross sections are considerably smaller, the strong bounds on resonances can be comparable or even stronger than the bounds from missing energy searches.
In this model, only bosons can couple directly to a pair of SM states, giving rise to potential resonances in two fermions or two gauge bosons or Higgses.
Fermions can potentially couple to a SM Higgs (or massive gauge boson) and a SM fermion, however such coupling will be proportional to the Yukawa coupling, thus it is numerically relevant only for third generation quarks.
Furthermore, resonant final states with a Higgs or massive gauge boson plus a quark are very difficult to reconstruct, therefore we will ignore this possibility in the following.
The interesting final states are:
\begin{center}
\begin{tabular}{lc}
 resonant final state &  mother particles in\\
    &  tiers $(2,0)$ and $(0,2)$ \\
\hline
$l^{+} \nu$  ($l^- \bar{\nu}$)& $W_{(2)}^\pm$ \\
$l^+ l^-$ & $A_{(2)}$, $Z_{(2)}$ \\
\hline
$t \bar{b}$ ($\bar{t} b$) &  $W^\pm_{(2)}$, $S^\pm_{(2)}$ \\
$t \bar{t}$ & $A_{(2)}$, $Z_{(2)}$, $G_{(2)}$, $H_{(2)}$, $S_{(2)}^0$ \\
$j j$ & $W^+_{(2)}$, $A_{(2)}$, $Z_{(2)}$, $G_{(2)}$ \\
\hline
$W^\pm Z$  &  $W_{(2)}^\pm$ \\
$W^+ W^-$   &  $A_{(2)}$, $Z_{(2)}$ \\
\hline 
 \end{tabular} \end{center} 
There exists searches looking for resonances in all these final states, however the most sensitive ones are the leptonic searches for $Z' \to l^+ l^-$ and $W' \to l^+ \nu$ resonances.
This holds true in this model too, even though the branching ratios into leptons are very small.
Jets, which have much larger recurrence, need to fight against a very large QCD background: searches of dijet resonances are only sensitive to rather large invariant masses (above $1$ TeV) and order pb cross sections.
On the other hand, there is an interesting CMS search for a pair of dijet resonances which will give interesting results in this case.
Resonances involving gauge bosons or tops pay the price of leptonic branching ratios for the $W$ and/or $Z$ decays, and severe cuts to reconstruct the heavy states, thus they will not be competitive.
In the following, we will try to estimate the reach of the relevant searches without performing a complete simulation but just counting the effective cross section for final states with the resonance and assuming the same experimental acceptance as for the model analysed in the experimental paper.
The limitations of this approach will be discussed case by case.

\subsection{Simulation details}

For the numerical study we make use of the Madgraph and CalcHEP interface of the model via FeynRules, described in the previous section.
All the cross sections used here were computed using Madgraph, and checked with CalcHEP.
We set the two free parameters of the model to $\Lambda R = 10$ and $m_{b} = 0$ GeV: the former controls the loop induced mass splitting and we fix it to the maximum value we would expect from naive dimensional analysis in 6D, the latter controls the mass of the Higgs resonances.
The Higgs resonances turn out to be irrelevant for the channels we are interested in here, thus changing $m_b$ will not affect our results significantly.
In all the calculations we consider both the symmetric radii case ($R_4 = R_5$, or $\xi=1$) and the asymmetric radii case ($R_4 \gg R_5$, or $\xi \gg 1$), as in the simplified FeynRules implementation.

\begin{figure}[tb]
\begin{center}
\includegraphics[width=7cm]{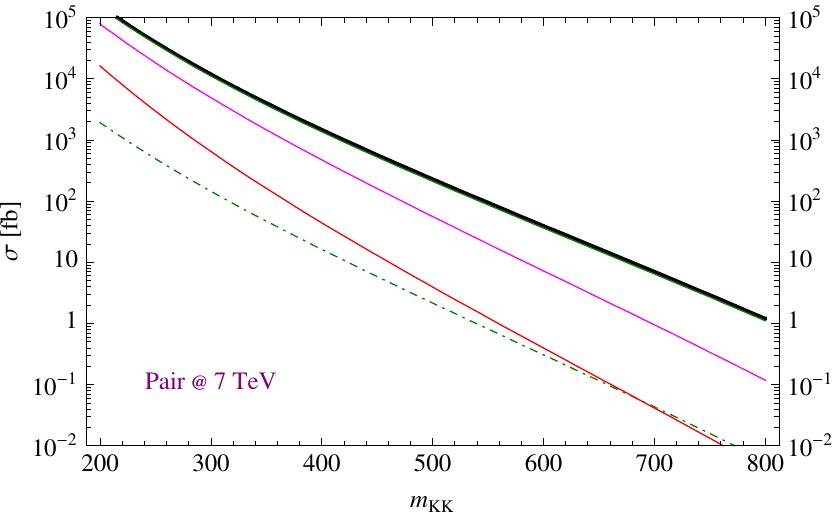} \includegraphics[width=7cm]{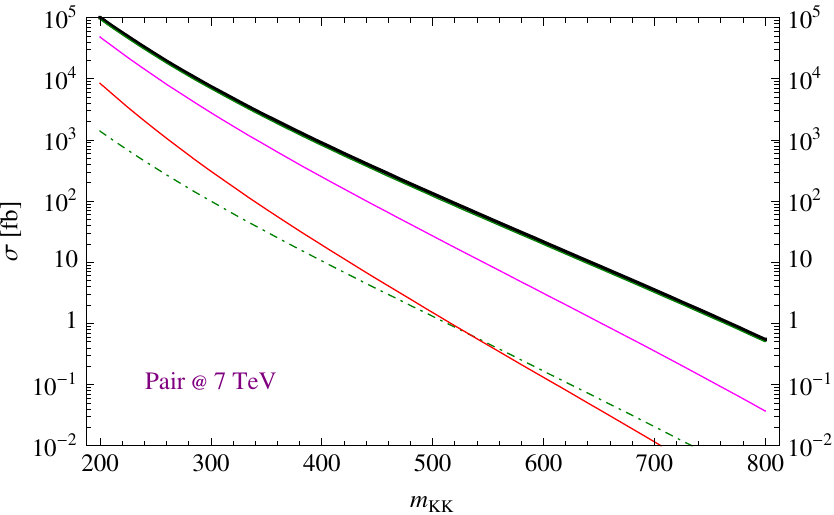}
\includegraphics[width=7cm]{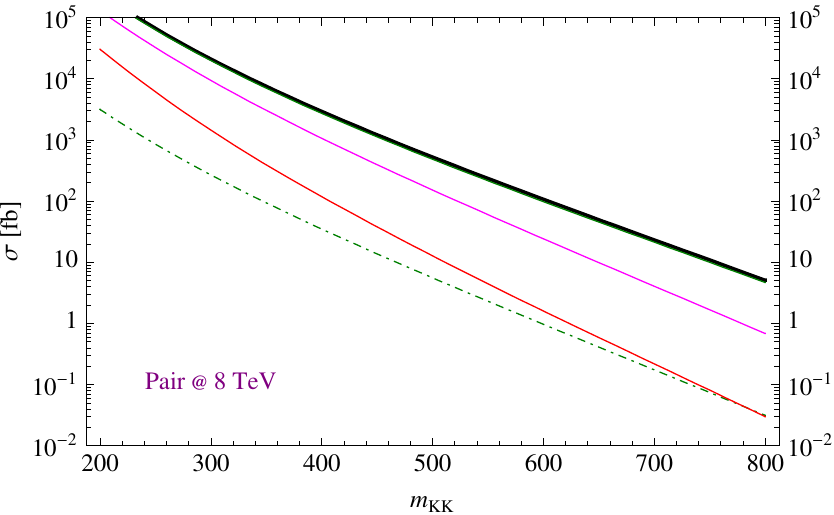} \includegraphics[width=7cm]{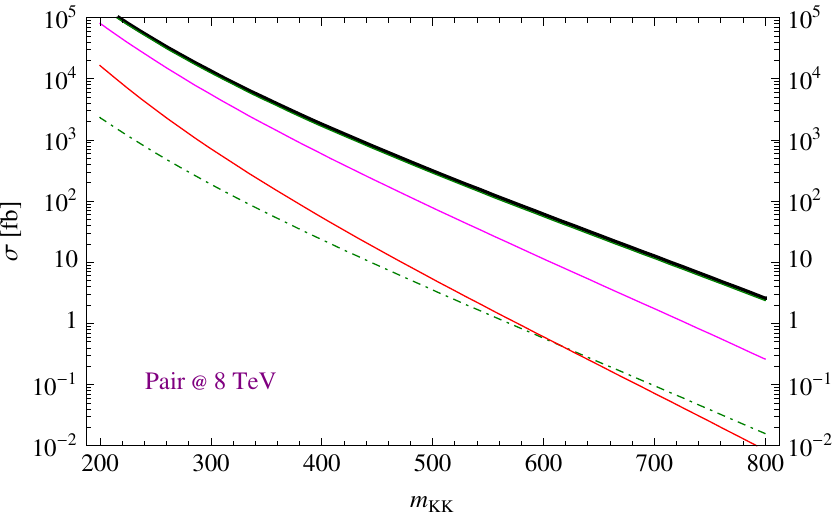}
\end{center}
\caption{\footnotesize Pair production cross sections of even heavy states for asymmetric (left two plots) and symmetric (right two plots) radii. The colours correspond to: total (black), $QQ$ (solid green), $GQ$ (magenta), $GG$ (red), and $QQ3$ (dashed green). The solid green line is barely visible, as it almost coincides with the black one.} \label{fig:xsectPair}
\end{figure}

\begin{figure}[tb]
\begin{center}
\includegraphics[width=7cm]{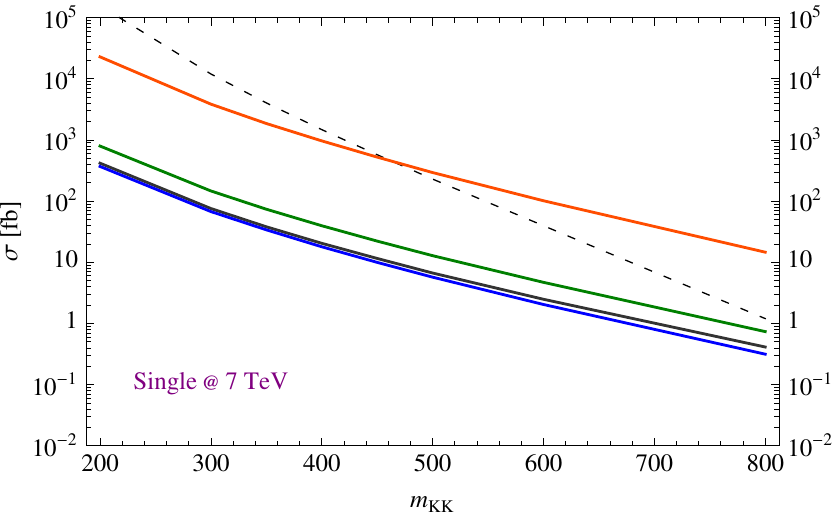} \includegraphics[width=7cm]{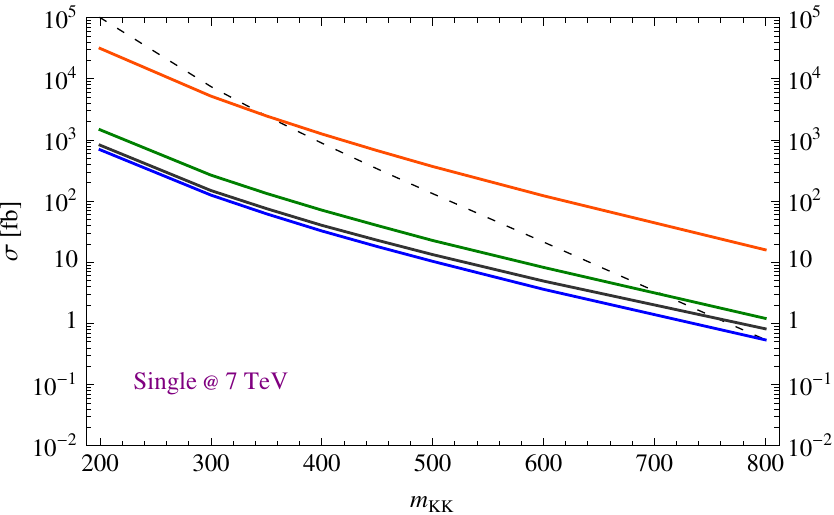}
\includegraphics[width=7cm]{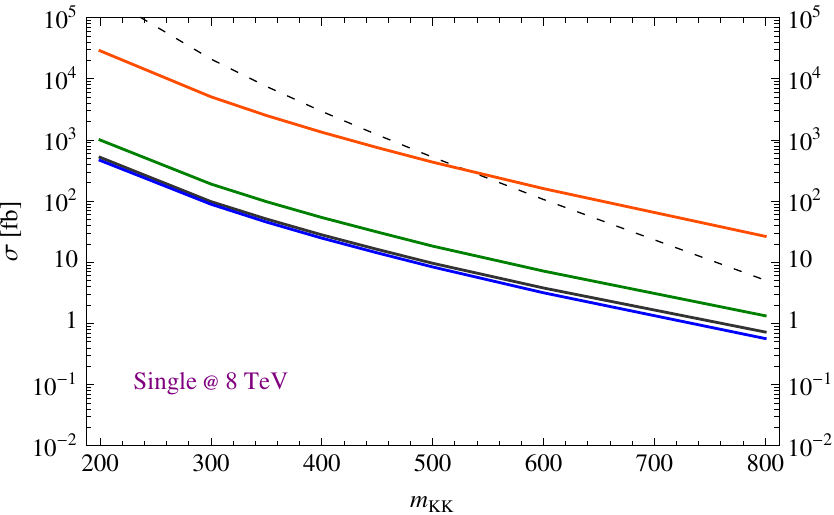} \includegraphics[width=7cm]{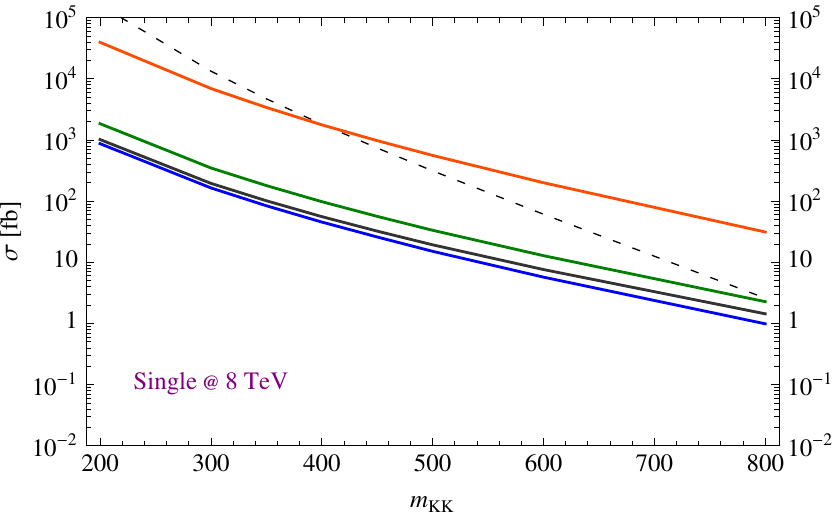}
\end{center}
\caption{\footnotesize Single production cross sections of even heavy states for asymmetric (left two plots) and symmetric (right two plots) radii. The colours correspond to: gluon $G_{(2)}$ (red), $W_{(2)}$ (green), $Z_{(2)}$ (blue) and $A_{(2)}$ (grey). For comparison, the dotted line is the total pair cross section.} \label{fig:xsectSing}
\end{figure}

\begin{figure}[tb]
\begin{center}
\includegraphics[width=7cm]{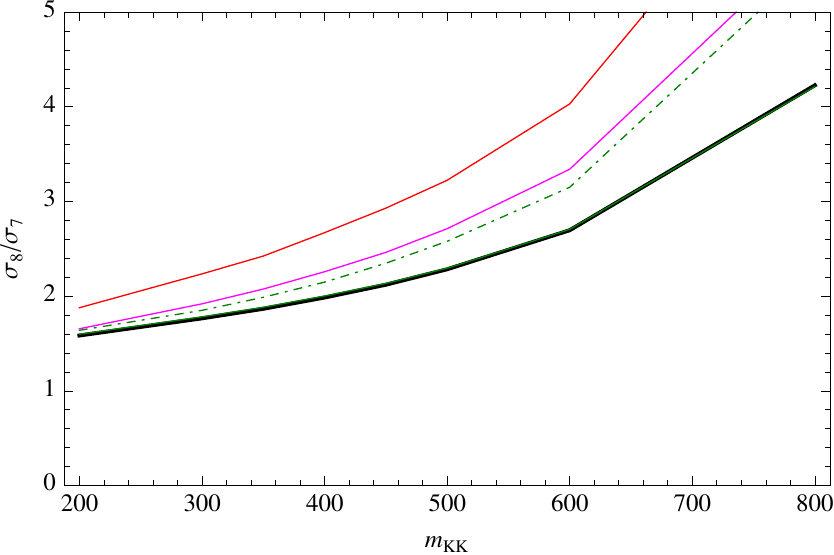} \includegraphics[width=7cm]{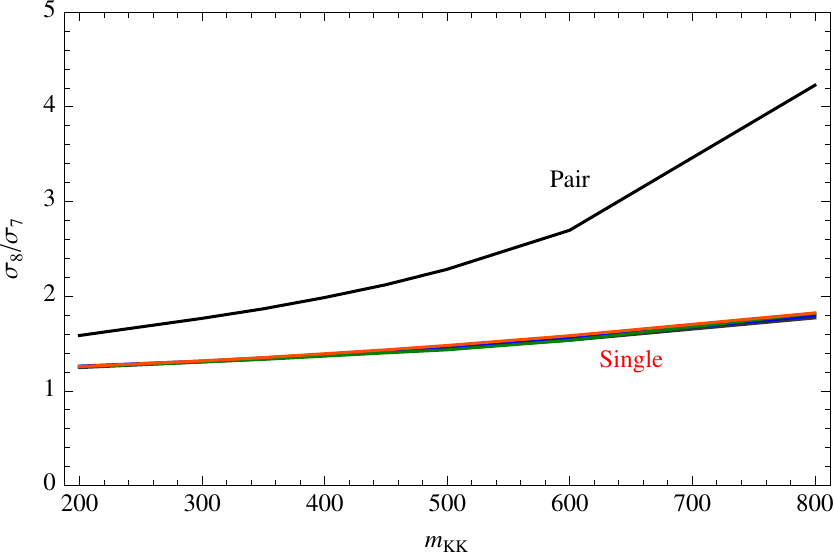}
\end{center}
\caption{\footnotesize Ratio of the $8$ TeV to the $7$ TeV cross section for pair production (left) and single production (right) channels. The same color codes as in the previous figures is adopted. The shown plots refer to the asymmetric case, similar values are obtained in the symmetric one.} \label{fig:xsectRatios}
\end{figure}

The resonant channels only come from electroweak gauge bosons $A_{(2)}$, $Z_{(2)}$ and $W^\pm_{(2)}$ or the gluon $G_{(2)}$.
The latter is the heaviest state in the tier, thus it can only be directly produced, while the electroweak bosons can derive form the decays of heavier states (the quarks).
Direct pair production of electroweak bosons is largely overwhelmed by the production via decays of coloured states, therefore we will only focus on the latter and neglect subdominant electroweak processes.
On the other hand, it is possible to singly produce all the gauge bosons via the loop induced coupling to the quarks: the resonant states can thus come from resonant Drell Yan production or from a chain decay of the gauge boson.
We will consider the following production channels:
\begin{itemize}
\item[-] Pair production:
              \begin{enumerate}
              \item[a)] pair of first and second generation quarks $QQ$ , including the processes $q_{(2)} \bar{q}_{(2)}$, $q_{(2)} q_{(2)}$ and $\bar{q}_{(2)} \bar{q}_{(2)}$ with all the flavour combinations;
              \item[b)] production of a quark in association with a gluon $GQ$, where the heavy quark is a first or second generation partner;
              \item[c)] pair production of gluons $GG$;
              \item[d)] pair production of third generation quarks $QQ3$, including $b_{(2)} \bar{b}_{(2)}$ and $t_{(2)} \bar{t}_{(2)}$.
              \end{enumerate}
\item[-] Single production:
              \begin{enumerate}
              \item[a)] single production of a gluon $G_{(2)}$, which can decay into leptons via a chain decay involving heavy quarks;
              \item[b)] single production of $Z_{(2)}$ and $W^\pm_{(2)}$, which can decay into leptons (or jets) either directly or via heavy leptons;
              \item[c)] single production of a $A_{(2)}$, which can only decay directly into a pair of leptons (or jets).
              \end{enumerate}
\end{itemize}
All those channels can produce resonant SM particle pairs at the end of a short or long decay chain.
Some examples of decay chains, with a resonant $\mu^+ \mu^-$ pair, are:
\begin{itemize}
\item[*]   $q^{(2)}_S \to q A^{(2)} \to q\, [\mu^+ \mu^-]$;
\item[*]   $q^{(2)}_D \to q' W^{(2)+} \to q' \nu e^{(2)+}_D \to q' \nu e^+ A^{(2)} \to q' \nu e^+\, [\mu^+ \mu^-]$;
\item[*]   $G^{(2)} \to q \bar{q}^{(2)}_D \to q \bar{q} Z^{(2)} \to q \bar{q}\, [\mu^+ \mu^-]$;
\item[*]   $Z^{(2)} \to \mu^+ \mu^{(2)-}_D \to \mu^+ \mu^- A^{(2)} \to \mu^+ \mu^-\,  [\mu^+ \mu^-]$;
\end{itemize}
where the two resonant leptons are in brackets.
The decay chain can contain quite a few additional SM particles, including jets and leptons, which can in principle complicate the reconstruction of the resonant pair. 
However, the small mass difference between states in the same tier will ensure that the momentum carried by the intermediate products in the chain is much smaller than the momenta of the resonant pair, which inherits the mass of the heavy gauge boson.
Therefore, it is a reasonable assumption that the soft intermediate decay products will not affect significantly the acceptance of the experimental cuts.
We will base our analysis on this assumption, leaving a detailed simulation for future work.
We also limit ourselves to leading order cross sections, even though QCD corrections may give rise to significant increase (or decrease) of the cross sections.
The results for the total cross sections are shown in Figure~\ref{fig:xsectPair} and~\ref{fig:xsectSing}.
From Figure~\ref{fig:xsectPair} we can see that all pair production cross sections are dominated by the $QQ$ channel, and that they drop very quickly with increasing $m_{KK}$: this is due to the fact that two states of mass $\sim 2 * m_{KK}$ must be produced.
Furthermore, the symmetric radii case fares slightly smaller cross sections, due to the fact that the masses are larger than the asymmetric case because of larger loop contributions.
We also see a significant increase from $7$ TeV centre of mass energy to $8$ TeV.
From Figure~\ref{fig:xsectSing}, on the other hand, we can see that single production cross sections decrease much more slowly with increasing $m_{KK}$, until they dominate for large masses ($m_{KK} \gtrsim 450 \div 500$ GeV for asymmetric radii).
In the symmetric radii case, the single production cross section is larger: this is due to the larger loop induced coupling. Single production, therefore, tends to dominate over pair production for smaller radii ($m_{KK} \gtrsim 350 \div 400$ GeV).
We also see that the increase of cross section to $8$ TeV is not as significant as in the pair production case.
This fact is more clear in Figure~\ref{fig:xsectRatios}, where we plot the gain from $7$ to $8$ TeV in the asymmetric case.
For pair production, the gain can be significantly larger than 1 and is particularly effective for final states involving gluons, while for single production channels the gain is always limited.

The main difference between the asymmetric and symmetric radii cases so far is that in the symmetric case the single production cross sections are much more important.
Another fundamental difference appears in the decay rates: in the asymmetric case, decays into odd gluons are kinematically forbidden because the mass correction to the odd gluon is too large.
In the symmetric case, the mass corrections to even ${(2)}$ states are doubled. 
As a consequence, decays into odd gluons open up: this means that for even quarks and gluons, the channel $q^{(2)} \to G^{(1)} q^{(1)}$ and $G^{(2)} \to G^{(1)} G^{(1)}$ will significantly increase the decay rate into odd states.
Therefore, rates into resonant pairs will be significantly reduced, particularly so in channels involving heavy quarks.

\subsection{$Z' \to l^+ l^-$ searches}

The most promising search is for a resonance in the dilepton channel, which is one of the cleanest channels at hadronic colliders like the LHC.
Both CMS~\cite{CMSll7} and ATLAS~\cite{ATLASll7} have published the final analysis of the full 2011 dataset at $7$ TeV, while only CMS has published preliminary results with the $8$ TeV run data~\cite{CMSll8}.
Both experiments search for final states with at least two energetic electrons or muons with opposite sign, and reconstruct the invariant mass of the pair.
The electron channel is slightly more efficient than the muon one, because the total energy can be measured in the electromagnetic calorimeter.
On the other hand, the muon momentum can only be reconstructed by measuring the curvature of the track in the muon chambers, and this procedure becomes more difficult for energetic muons.
The expected signal is a Gaussian bump in the dilepton invariant mass distribution, and the efficiency of the signal is estimated by simulating a narrow resonance for which the width of the Gaussian peak is determined by the experimental resolution on the lepton momentum reconstruction.

\begin{figure}[tb]
\begin{center}
\includegraphics[width=7cm]{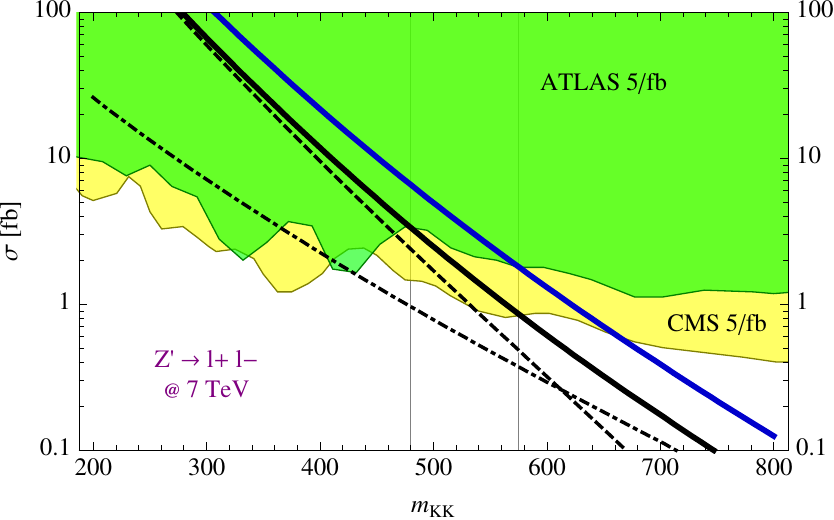} \includegraphics[width=7cm]{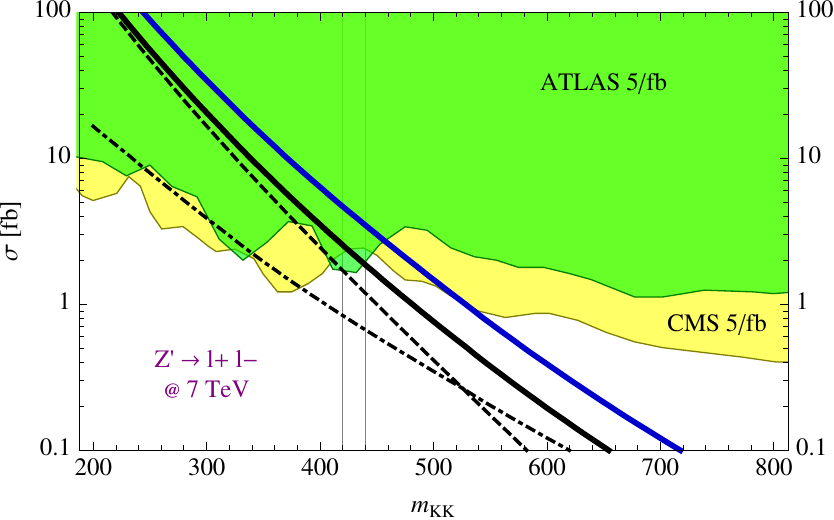}
\end{center}
\caption{\footnotesize Leptonic $Z'$ bounds in the asymmetric (left) and symmetric (right) cases. The solid black line represents the total cross section, while the dashed (dash-dotted) one shows the pair (single) production component. The blue line is, for comparison, the $8$ TeV total cross sections. The CMS (ATLAS) excluded region is shown in yellow (green). } \label{fig:boundsLL}
\end{figure}

In the model under investigation, the resonant dilepton pair is accompanied by extra soft leptons coming from the intermediate decay chain, however, the momenta of the resonant pair should be much larger than the soft intermediate ones.
Therefore, the presence of extra leptons should not affect the reconstruction of the peak.
Furthermore, the decaying bosons are very narrow, with widths of the order of 0.005\% of the mass for $A^{(2)}$ and 0.01\% for the $Z^{(2)}$.
The most serious issue is that the resonant lepton pairs come from two different states with different masses:  the loop corrections generate a mass splitting between the $Z^{(2)}$ and the $A^{(2)}$ which amounts to about $5\%$ of $m_{KK}$ ($\sim 2.5\%$ of the mass) in the asymmetric case and $9.5\%$ of $m_{KK}$ ($\sim 5\%$ of the mass) in the symmetric case.
The CMS search at $7$ TeV~\cite{CMSll7} reports that the resolution on the dimuon invariant mass is 6.5\% at masses around $1$ TeV, increasing to 12\% at $2$ TeV, while the resolution in the dielectron channel is 1.1\% for electron in the barrel and 2.3\% when one electron is in the endcap calorimeter, staying constant above $500$ GeV.
These numbers show that the electron channel has enough resolution to disentangle the two resonances, and a dedicated search may give very tight bounds on this model.
For the muon channel, the experimental resolution is still too large.
The experimental bounds on a $Z'$ mass is performed by using an unbinned likelihood analysis, where the signal is modelled by a Breit-Wigner function convoluted with a Gaussian resolution function.
The width in the Breit-Wigner function is set to be equal to a sequential $Z'$, thus amounting to $3.1\%$ of the mass.
The single bump used in setting the bounds would in all cases encompass both resonances in this model, therefore in the following we will assume that both resonances contribute to the same bump and sum the two cross sections when comparing the expected cross section to the experimental bound.
A similar strategy is used by the ATLAS collaboration~\cite{ATLASll7}: for their detector, the energy resolution on electrons is estimated to be 1.2\% in the barrel and 1.8\% in the endcap regions, while for muons the $p_T$ resolution ranges from 10\% to 25\%.

In this model, the effective branching ratios into a resonant pair of leptons are very small: they range from 0.2\% to 0.5\% for coloured states in the asymmetric radii case, and from 0.05\% to 0.3\% in the symmetric case.
Nevertheless, the total cross section is still quite large and comparable to the experimental bounds.
The results of our calculation are shown in Figure~\ref{fig:boundsLL}, where the black lines show the total cross section while the dashed (dot-dashed) ones show the component from pair (single) production channels only.
The experimental bounds, in green for ATLAS and yellow for CMS, refer to the combined results with 5 fb$^{-1}$ of data at $7$ TeV, where we can see that at large masses the CMS bound prevails.
In the asymmetric case, the best bound is at $m_{KK} \sim 575$ GeV, which is in a mass region where the pair and single production channels are comparable.
In the symmetric case (right plot), we see that the pair production channel is highly suppressed because of the heavier mass, while the single channel is suppressed to a smaller degree due to the smaller branching ratios.
The net result is a bound at around $m_{KK} \sim 440$ GeV, dominated by pair production.

The CMS collaboration has also published preliminary results on an analysis performed on the 2012 data at 8 TeV, however no combined electron-muon results is reported.
The best quoted bound comes from a combination of the 7 and 8 TeV datasets, which is done by rescaling the signal at 7 TeV to 8 TeV.
As this rescaling is model dependent, we cannot directly use this result here.
Nevertheless, in the Figure~\ref{fig:boundsLL}, we show for comparison the expected cross section at 8 TeV in blue

\subsection{$W' \to l \nu$ searches}

\begin{figure}[tb]
\begin{center}
\includegraphics[width=7cm]{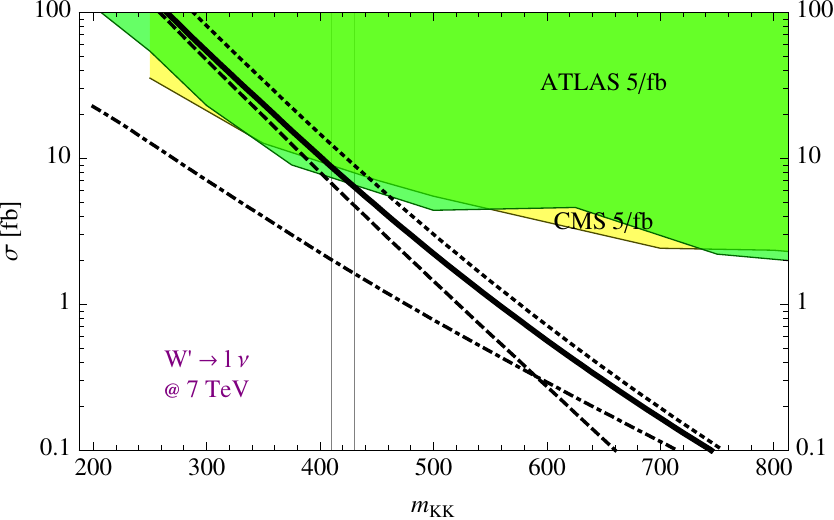} \includegraphics[width=7cm]{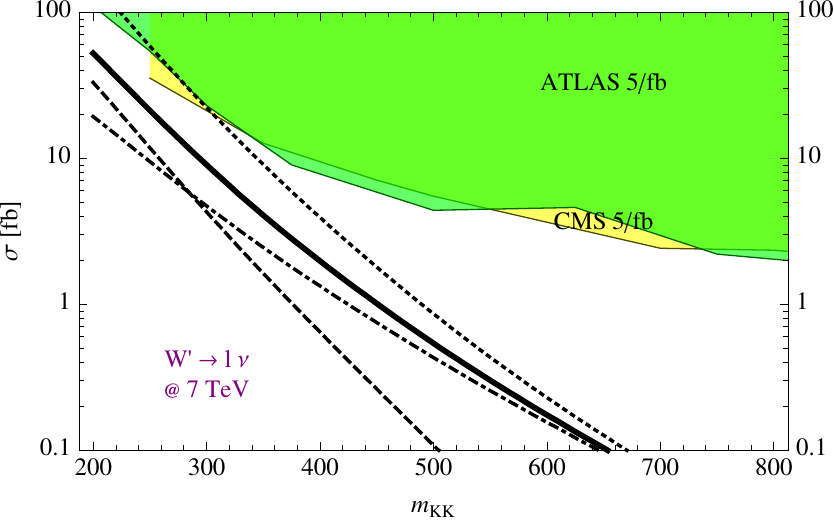}
\includegraphics[width=7cm]{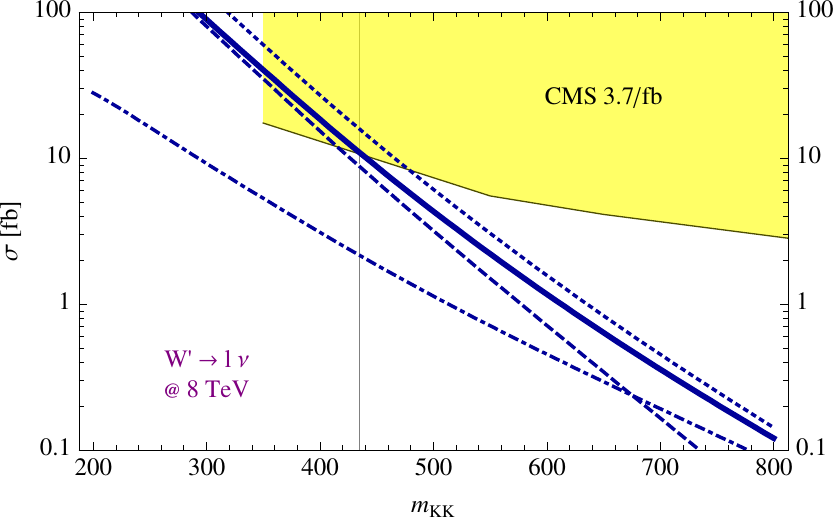} \includegraphics[width=7cm]{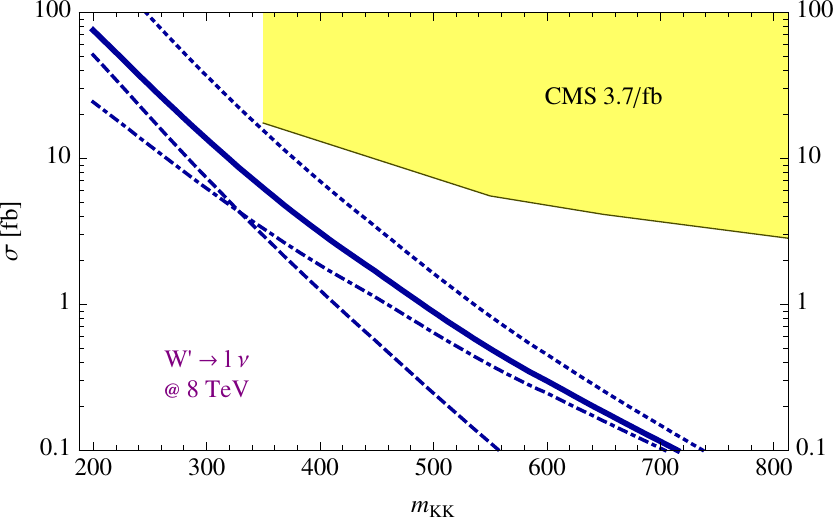}
\end{center}
\caption{\footnotesize Leptonic $W'$ bounds in the asymmetric (left) and symmetric (right) cases. The solid lines represent the total cross section, while the dashed (dash-dotted) one shows the pair (single) production component. The dotted line is the total cross section without the MET veto. The upper plots refer to the $7$ TeV data, while the lower ones to the $8$ TeV CMS analysis. The CMS (ATLAS) excluded region is shown in yellow (green). } \label{fig:boundsLN}
\end{figure}

The second most promising analysis is the search for a $W'$ in the leptonic channel.
Both CMS~\cite{CMSln7} and ATLAS~\cite{ATLASln7} collaborations have published their final searches with the 2011 7 TeV dataset, while only CMS~\cite{CMSln8} has preliminary results available for the 2012 8 TeV dataset.
This search is based on the requirement of a single high transverse momentum $p_T$ electron or muon in association with missing transverse momentum (which may derive from the neutrino).
The main observable used to isolate the signal is the transverse mass $M_T$, defined as
\begin{equation}
M_T = \sqrt{2 \cdot p_T^l \cdot p_T^{\rm miss} \cdot (1-\cos \Delta \phi)}\,,
\end{equation}
where $p_T^l$ is the transverse momentum of the electron, $p_T^{\rm miss}$ is the missing transverse momentum in the event and $\delta \phi$ is the azimuthal angle between the lepton and missing transverse momenta.
A signal from a massive $W' \to l \nu$ would produce a Jacobian peak in the $M_T$ distribution at the $W'$ mass, therefore a lower cut on $M_T$ is imposed somewhat below the $W'$ mass used in the simulated signal.
As before, the presence of intermediate decay products should not significantly affect this search.
However, in the pair production channels, the decays of the second heavy state may contain additional invisible particles which will contribute to the total $p_T^{\rm miss}$ of the event and distort the $M_T$ distribution.
The most worrisome cases involve the decay to a pair of odd states, which have significantly large branching rations.
To be conservative, in this analysis we will veto on the decay of the second heavy state into an odd pair, nevertheless the result without this veto are also shown.
The actual bound should lay in between these two lines.

Both CMS and ATLAS extract the bounds on a $W' \to l \nu$ that does not interfere with the SM, and for each mass point analysed they impose an optimised cut on $M_T$.
Here for simplicity we will assume that the efficiency of the experimental cuts are the same as the ones in the model they consider, and we simply compare the value of our effective cross section to the bound.
The results are shown in Figure~\ref{fig:boundsLN} for both asymmetric and symmetric radii, and for 7 and 8 TeV studies. At 7 TeV, the bounds from ATLAS and CMS are comparable.
In the asymmetric case, the bound is significantly lower than the the dilepton channels, and it amounts to $m_{KK} \sim 430$ ($435$) GeV for the 7 (8) TeV analysis.
The bounds is in a region dominated by the pair production, and the inclusion of events with a pair of odd states would improve the bound by about $50$ GeV.
In the asymmetric case, as before, we can see a significant decrease of the pair production channels and a less important decrease in the single production.
The decrease is however sufficient to remove the bound, as the predicted cross section always lies below the excluded region.
The inclusion of events with an odd pair would significantly enhance the event rate, with a maximum bound of $m_{KK} \sim 300$ GeV.

\subsection{Paired dijet search}

The coloured heavy states have rather large branching ratios in events with a resonant jet pair, ranging from 35\% to 65\% in the asymmetric case, and from 12\% to 23\% in the symmetric case.
However, analyses based on jets suffer from large QCD background which are difficult to deal with.
Searches for dijet resonances have been published from both ATLAS and CMS, however the bound on the cross section is significant only in regions with very high invariant mass (above a TeV).
On the other hand, CMS has published an interesting search of events with a pair of dijet resonances with similar mass~\cite{CMSjjjj}, based on 2.2 fb$^{-1}$ integrated luminosity from the 2011 data.
Due to the large branching ratios, the pair production channels are due to produce a fairly large number of such states, therefore this search may be competitive for this model.

The search strategy is the following: the selected events must have at least 4 jets with transverse momentum $p_T > 150$ GeV.
The four highest $p_T$ jets are combined into two pairs with a large separation between the two pairs in order to minimise overlaps, and, out of all the combinations, the search selects the one with the largest $\Delta m/m_{avg}$, where $\Delta m$ is the difference in invariant mass and $m_{avg}$ is the average invariant mass of the two pairs.
Finally, only events with $\Delta m/m_{avg} < 15\%$ are selected in order to minimise the QCD multijet background.
An additional cut on $\delta = p_{T, pair 1} + p_{T, pair 2} - m_{avg} > 25$ GeV is also imposed to further reduce the QCD background and have a smooth tail of background events at $m_{avg} > 320$ GeV.
These selections do not suffer from the presence of the soft intermediate products, however a significant factor in the game is the fact that the dijet pairs in this model may arise from 4 states with different masses: $A^{(2)}$, $Z^{(2)}$, $W^{(2)}$ and $G^{(2)}$.
The mass slitting between the $Z^{(2)}$ and $W^{(2)}$ with the lighter $A^{(2)}$ is of the order of 2.5\% (5\%) of the mass in the asymmetric (symmetric) case, while the splitting of the gluon $G^{(2)}$ with $A^{(2)}$ is of the order of  10\% (18\%).
While the splitting between electroweak gauge bosons is small compared to the $15\%$ cut on $\Delta m/m_{avg}$, events with gluons may fall outside of the selected region.
This problem is not severe, though, because the signal is dominated by the electroweak gauge bosons: more than 80\% of the signal, therefore the signal loss would be limited.
The reason for this is that the $G^{(2)}$ is the heaviest particle in the tier, therefore it cannot appear in decay chains of lighter states like quarks.
Thus, the signal with a $G^{(2)}$ resonance paired with an electroweak boson can only appear in the GQ or GG channels, with one gluon decaying directly into a pair of jets.
In the following we will neglect this potential signal loss, and consider the total cross section.

\begin{figure}[tb]
\begin{center}
\includegraphics[width=7cm]{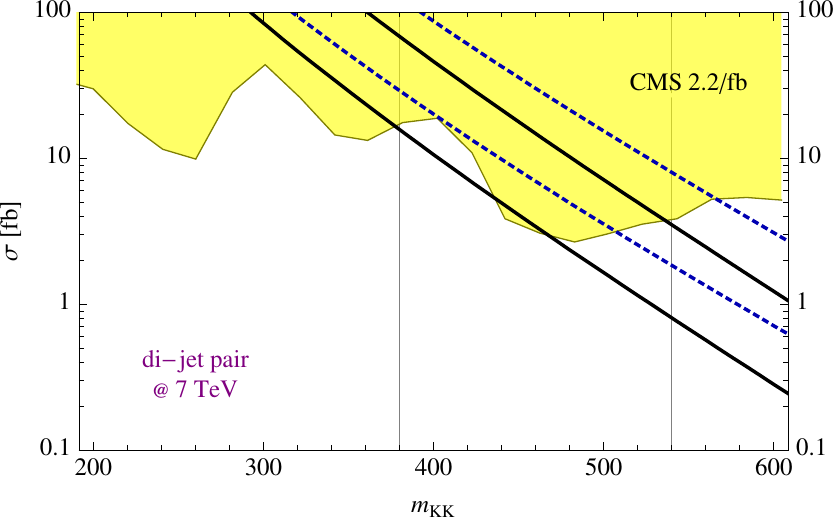} \includegraphics[width=7cm]{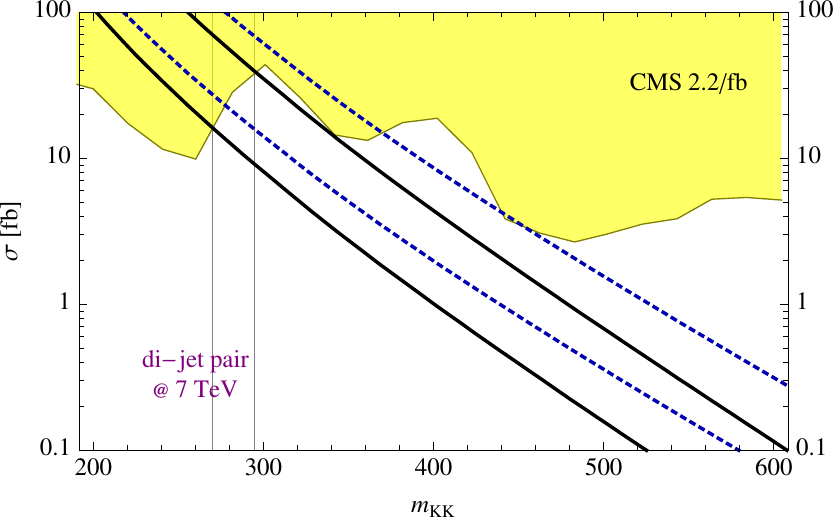}
\end{center}
\caption{\footnotesize Di-jet pair bounds in the asymmetric (left) and symmetric (right) cases. The solid black lines show the total effective cross section assuming an experimental efficiency of 3\% and 13\% respectively. The dashed blue lines are, for comparison, the $8$ TeV total cross sections. The CMS  excluded region is shown in yellow.} \label{fig:bounds4j}
\end{figure}

Another important issue in this channel is the fact that the published bound is on the production cross section times branching ratio times the experimental acceptance.
The model considered in the CMS search is a model with scalar colour octets (coloron), which can decay into a pair of quarks $q \bar{q}$, and they are pair produced via QCD interactions.
The experimental acceptance for this model is 3\% for a coloron mass of $300$ GeV, raising to 13\% for a mass of $1$ TeV.
The search has been performed for masses in the range $320$ to $1200$ GeV.
To extract a reliable bound from this search, a detailed simulation of the signal is indispensable, in order to extract the experimental acceptance.
However, in order to have an estimate of the bound, we will here assume that the acceptance is within 3\% to 13\%, and leave a detailed simulation for future work.

The result of our calculations are shown in Figure~\ref{fig:bounds4j}, where the two black lines correspond to the two coloron acceptances, and the CMS excluded region in yellow is based on a 7 TeV dataset of 2.2 fb$^{-1}$.
The curves show that in the asymmetric case, this search is potentially relevant due to the fairly large cross sections obtained in this case.
Assuming the coloron acceptance, the bound on $m_{KK}$ is expected to range from $380$ to $540$ GeV, a region which is close to the dilepton channel.
Moreover, this bound is only based on half of the total 2011 dataset.
The blue dashed lines in the plot show the predicted cross section at 8 TeV.
In the symmetric case, the prospect is much more difficult: in fact, the cross section is heavily suppressed because it only receives contribution from the pair production channels, and we expect a bound in the $300$ GeV mass region, thus well below the dilepton channel.

\section{Conclusions}
\label{sec:concl}

The excellent performance of the LHC and of the related experiments has allowed to collect a remarkable amount of high quality data, 5 fb$^{-1}$ at 7 TeV in 2011 and an expected 30 fb$^{-1}$ at 8 TeV by the end of 2012.
This has allowed for the discovery of a new state that resembles the Standard Model Higgs boson, and to place very strong bounds on the masses of constrained supersymmetric models and other simple extensions of the SM.
The experimental results, which does not show any sign of physics beyond the predictions of the SM, however, are yet not sufficient to disfavour the naturalness argument which would prefer new states at the TeV scale.
In fact, more work and data are necessary to strongly constraint top partners or alternative models to supersymmetry.
In this paper we focused on one alternative scenario, where the Dark Matter particle that fills the Universe is provided by extra dimensions.
In fact, symmetries of the compact space, relic of the extended Poincar\'e invariance after the compactification, can ensure the stability of the lightest Kaluza-Klein resonance of the SM states.
The presence of such a symmetry is indeed a remarkably strong requirement, allowing for instance to select a single flat orbifold in 5 or 6 dimensional models: the Real Projective plane (RP$^2$).

\begin{table}[t]
\centering
\begin{tabular}{|l|c|c|}
\hline
Channel & asymmetric & symmetric \\
\hline
$Z' \to l^+ l^-$ & $575$ & $440$ \\
$W' \to l \nu$  & $430$ & $-$ \\
paired dijet & $380 \div 540$ & $270\div 295$\\
\hline
\end{tabular}
\caption{\footnotesize Estimated bounds on $m_{KK}$ in the asymmetric ($\xi \gg 1$) and symmetric ($\xi = 1$) cases.
}
\label{tab:summary}
\end{table}

In this paper we study a UED realisation of the RP$^2$, where a 6D field is associated to each SM field, including the Higgs field.
Loop corrections are essential to study the phenomenology of this class of models.
In this work, we completed the one loop analysis of the first odd and even tiers, by presenting the complete calculation of the one-loop corrections to the masses of the even tiers, and we generalised the results for the odd tiers~\cite{Cacciapaglia:2009pa} to the case of asymmetric radii.
The results for the loop induced couplings of the even tiers can be found in~\cite{Cacciapaglia:2011hx}.
These loop results are essential to study the phenomenology of the model, as they determine the decay and production rates of the new states.
In this work, we estimated the bounds from resonant decays of even gauge bosons into a pair of standard model particles.
We identified 3 promising channels: dilepton resonances, decays into a single lepton and neutrino, and decay into dijet of a pair of states.
In the estimate, we simply count the effective cross section of the final states of interest, and we assume that the experimental efficiency is the same as the model used by the experimental groups to publish the bounds.
This procedure is approximate, and a detailed simulation would be necessary to extract more reliable numbers; nevertheless, the estimates can give us an idea of the relevance of the bounds.
The results are summarised in the Table~\ref{tab:summary}: the most promising channel is the dilepton resonance, that gives a lower bound on $m_{KK}$ of $575$ ($440$) GeV in the asymmetric (degenerate) case.
However, to extract the estimate we assumed that all the events fall under the same Gaussian peak, while in reality they arise from two separate peaks from $Z_{(2)}$ and $A_{(2)}$.
A fit to the data that takes into account the presence of two peaks may improve the bound, especially in the electron channel that has better energy resolution.
A surprisingly competitive channel is the search for paired dijet resonances: in this case, the bound depends crucially on the efficiency of the selection rules. Furthermore, the CMS study is only based on 2.2 fb$^{-1}$ of data at 7 TeV, thus the bound may be significantly improved if the QCD background is properly taken into account.
Our estimate, based on the efficiencies quoted on the CMS paper and based on a scalar coloron model, show that in both cases this search may be competitive with the dilepton one.
Other channels, like $t \bar{t}$ resonances, $t b$ resonances, $W V$ ones and dijet, are not competitive as they are mostly sensitive to large invariant masses and large cross section. The price to be paid here is either a small leptonic branching ratio of the $W$ and $Z$, or the large QCD background.
The resonant channels are very important because they simplify the searches, and also because such signatures are not present in supersymmetric models.
The UED model also have more supersymmetry-like signatures coming from the chain decays of the odd tiers, ending into a stable neutral particle.
In this case, the signal will resemble that of a compressed supersymmetric spectrum, and the most sensitive channel should be the one with jets and missing transverse momentum.
A detailed analysis of the reach of this channel is in progress~\cite{jetmet}.

Contrary to supersymmetry, the masses of the KK resonances cannot be pushed to an arbitrary large mass scale: in fact, the calculation of the relic abundance selects a finite range for the preferred value of $m_{KK}$, and the overclosure of the Universe sets an upper bound.
Even considering the uncertainties related to our knowledge of the Cosmological model, this feature persists.
In Ref.~\cite{Cacciapaglia:2009pa} the preferred range has been estimated using an analytical approximation and only considering co-annihilation of all the states in the odd tier: the result is a preferred range of $300 < m_{KK} < 400$  ($210 < m_{KK} < 280$) for asymmetric (degenerate) radii.
However, this rough estimate is not enough, as it does neglect the impact of the even tiers~\cite{kakizaki}, that tend to push the preferred range towards higher values.
A detailed analysis of these effects will be presented in a forthcoming publication~\cite{RP2DM}.

The RP$^2$ also offer other phenomenologically interesting aspects, that have not been explored yet.
For instance, the odd tiers $(2,1)$ and $(1,2)$ have masses close to the even ones under study, $m_{(2,1)} = \sqrt{5} m_{KK} \sim 2.25\; m_{KK}$, thus their production rates will be similar.
Furthermore, they contain twice as many fermions.
The only allowed decay modes are decays within the tier, or decays to a lighter odd state plus a SM one, $(2,1) \to (0,1) + (0,0)$, via loop induced couplings.
These tiers will therefore be a novel source of events with jets and missing transverse momentum, and they may significantly affect the bound on the radius.
There are also tiers that can only decay via localised counterterms, like $(1,1)$ or the $(2)_-$ in the degenerate case: they may give rise to resonances or new sources of missing energy depending on the localised interactions.

\section*{Acknowledgments}
We thank J\'er\'emie Llodra-Perez for useful discussions, comments and for initiating the FeynRules implementation of the model.
The initial stages of the work of B.K. has been supported by the UNIVERSENET Marie Curie Research Training Network (contract n. MRTN-CT-2006-035863).

\appendix

\section{Appendix: loop functions}
\label{app:functions}
\setcounter{equation}{0}

The contribution of the torus to the loops is proportional to the function
\beq
T_6 (\xi) &=& \frac{1}{\pi \xi} \sum_{ \tiny \begin{array}{c}
n_1, n_2 \subset Z \\
\{n_1, n_2\} \neq \{0,0\} \end{array}} \frac{1}{(n_1^2 + n_2^2/\xi^2)^2}\\
&\sim& \frac{\pi^2}{45 \xi} + \xi^2 \left( \zeta(3) + \coth \left( \frac{\pi}{\xi} \right)-1\right) + \frac{\pi \xi}{\sinh^2 \left(\frac{\pi}{\xi}\right)}\,.
\eeq
The approximation is very good for moderately large values of $\xi$, i.e. for $\xi \lesssim 5$.
Note that
\beq
\xi^2 T_6 (1/\xi) = T_6 (\xi)\,,
\eeq
thus the torus contribution to $(n,0)$ and $(0,n)$ modes is the same.
$T_6 (\xi)$ is a rapidly increasing function of $\xi$: for instance, $T_6 (1) \sim 1.92$ while $T_6 (2) \sim 6.7$.

The KK number dependent functions generated by the glides, can be expressed in terms of the following integrals:
\beq
\Phi_1 (n) &=& 2 \pi^3 \int_0^\infty dk\; \frac{k^3}{\sqrt{k^2+n^2} \sinh \pi \sqrt{k^2+n^2}}\,, \\
\Phi_2 (n) &=& 2 \pi^3 \int_0^\infty dk\; \frac{k n (\sqrt{k^2+n^2}-n)}{\sqrt{k^2+n^2} \sinh \pi \sqrt{k^2+n^2}}\,, \\
\Phi_3 (n) &=& 2 \pi^3 \int_0^\infty dk\; \frac{k^3 (\sqrt{k^2+n^2}-n)}{n \sqrt{k^2+n^2} \sinh \pi \sqrt{k^2+n^2}}\,.
\eeq
These integrals are exponentially suppressed for large $n$, thus the impact is reduced for higher modes.

The functions appearing in loops for gauge bosons and fermions read
\beq
S_g (x) &=& 2 \Phi_1 (x) - 16 \Phi_2 (x) - 8 \Phi_3 (x)\,, \\
S_s (x) &=& \Phi_1 (x) - 4 \Phi_2 (x) - 4 \Phi_3 (x)\,, \\
V_g (x) &=& \frac{2}{3} \Phi_1 (x) - \frac{16}{3} \Phi_2 (x) + \frac{8}{3} \Phi_3 (x)\,, \\
V_s (x) &=& \frac{1}{3} \Phi_1 (x) + \frac{4}{3} \Phi_2 (x) + \frac{4}{3} \Phi_3 (x)\,, \\
F_g (x) &=& - \Phi_1 (x) - 4 \Phi_2 (x)\,, \\
F_s (x) &=& - \frac{1}{2} \Phi_1 (x)\,.
\eeq

\end{document}